\documentclass[12pt,oneside,english]{amsart}
\usepackage[T1]{fontenc}
\usepackage[latin9]{inputenc}
\usepackage{geometry}
\usepackage{amsthm}
\usepackage{amstext}
\usepackage{amssymb}
\usepackage{graphicx}

\makeatletter

\newcommand{\lyxdot}{.}

\numberwithin{equation}{section}
\numberwithin{figure}{section}

\makeatother

\usepackage{babel}
\begin{document}

\title{Incubation period, anti-retroviral therapies, age-structures}

\maketitle
\textbf{Full Title: }Incubation Periods Under Various Anti-Retroviral
Therapies In Homogeneous Mixing And Age-Structured Dynamical Models:
A Theoretical Approach.

\begin{center}

\vspace{0.6cm}

\textbf{Arni S.R. Srinivasa Rao}%
\footnote{Current address: Georgia Regents University, 1120 15th Street, Augusta,
GA 30912, USA. Email: arrao@gru.edu. Part of the revision was done
when the author was working at Bayesian and Interdisciplinary Research
Unit (BIRU), Indian statistical Institute, Kolkata 700108, INDIA%
},%
\footnote{This paper was benefited by several useful comments from Philip Maini,
Masa Kakehashi, Kurapati Sudhakar, Thomas Kurien, Ramesh Bhat. Comments
for revision by handling Associate Editor were very helpful. Supported
by funds from the Institute of Public and Preventive Health, Georgia
Regents University, Augusta. My sincere gratitude to all. %
}

Mathematical Institute, 

Centre for Mathematical Biology, 

University of Oxford, 24-29 St Giles', 

Oxford, OX1 3LB, England. 

\end{center}

\vspace{0.2cm}

(Accepted in \emph{The Rocky Mountain Journal of Mathematics, USA)}

\vspace{0.4cm}
\begin{abstract}
With the launch of second line anti-retroviral therapy for HIV infected
individuals, there has been an increased expectation on surviving
period of people with HIV. We consider previously well-known models
in HIV epidemiology where the parameter for incubation period is used
as one of the important components to explain the dynamics of the
variables. Such models are extended here to explain the dynamics with
respect to a given therapy that prolongs life of an HIV infected individual.
A deconvolution method is demonstrated for estimation of parameters
in the situations when no-therapy and multiple therapies are given
to the infected population. The models and deconvolution method are
extended in order to study the impact of therapy in age-structured
populations. A generalization for a situation when $n-$types of therapies
are available is given. Models are demonstrated using hypothetical
data and sensitivity of the parameters are also computed.
\end{abstract}



\tableofcontents{}

\pagebreak
\begin{abstract}
With the launch of second line anti-retroviral therapy for HIV infected
individuals, there has been an increased expectation on surviving
period of people with HIV. We consider previously well-known models in epidemiology where the
parameter for incubation period is used as one of the important components
to explain the dynamics of the variables. Such models are extended
here to explain the dynamics with respect to a given therapy that
prolongs life of an HIV infected individual. A deconvolution method
is demonstrated for estimation of parameters in the situations when
no-therapy and multiple therapies are given to the infected population.
The models and deconvolution method are extended in order to study
the impact of therapy in age-structured populations. A generalization
for a situation when $n-$types of therapies are available is given.
Models are demonstrated using hypothetical data and sensitivity of
the parameters are also computed.
\end{abstract}

\section{\textbf{Preliminaries, Basic ODE Model and Integro-Differential Equations
models}}

With the introduction of second line therapy\cite{WHO2010} to the
people living with HIV who were already on first line therapy until
introduction of second line therapy, there is a further hope to increase
the active life of HIV infected individuals. Revised estimates of
the people living with HIV are obtained for some countries to address
the impact of second line therapy (see, for example\cite{RaoAMSNotices2012}).
Second line theory is provided after failure to responding to the
first line therapy among the infected individuals. Modeling the impact
of second line therapy and corresponding extended survival time is
complicated because susceptible population can acquire virus from
two infected class of populations who are on therapy in addition to
the infected population who are without any therapy. However identifying
the first line individuals who are no more responding to the first
line therapy through surveillance is still a challenging issue in
several countries. Difficulties in monitoring and recording HIV infected
population who are on first line and second line therapy will also
lead to difficulty in estimating parameters of disease progression
and disease related mortalities. Disease progression rate and incubation
period are related and usually both are taken as reciprocal to each
other. Incubation period of HIV infected individuals is also expected
to increase since new anti-retroviral therapy policies. The incubation
period is generally defined as `the time duration between the time
a virus or bacteria enters the human body and the time at which clinical
clinical symptoms occur'. This duration can vary from case to case
depending upon the route through which the virus or bacteria enters
the immune system of an individual and in some cases depends upon
the age of the infected individual. For chickenpox this duration is
10 - 21 days, for common cold 2 - 5 days, for mumps 12 to 25 days,
for SARS a maximum of up to 10 days, for rubella 14 - 21 days, for
pertussis 7 - 10 days, and for HIV infection to AIDS 6 months to 10
years or more. The incubation period can be used as a measure of rapidity
of the illness after interaction with the virus or bacteria. It is
not easy to collect information on the incubation period of infected
individuals unless they are monitored. One of the direct ways of estimating
the average incubation period of a given virus in the population is
by surveillance and followup of the infected individuals until they
develop symptoms. All the infected individuals may not be aware of
their infection until symptoms appear and followup is subject to the
availability of an individual. It might not be possible to follow
up individuals in a typical situation, where time taken for the onset
of symptoms from the infection is longer, especially if infected individuals
are lost to followup. Hence, there are limitations on directly estimating
the average incubation period from prospective cohort studies. Nevertheless,
the incubation period occupies an important role along with other
parameters in modeling the disease spread and understanding the basic
reproductive rate. A useful description of various epidemic models,
and of estimation of parameters like the incubation period, transmission
rates, forces of infections are presented in \cite{AAM91}. The degree
of importance of obtaining accurate average incubation periods varies
with the incubation period of the disease. This degree of variation
causes mathematical models to act sensitively in predicting future
burden. Models describing dynamics of disease spread where the incubation
period is shorter are less subject for producing misleading results
than models for the spread with longer and varying incubation periods.
Especially for predicting AIDS, the epidemic models developed, depend
heavily on parameters that determine transmission rates of infection
from infected to susceptible and on the parameter which explains the
average time to progress to AIDS. A review of various modeling approaches
and quantitative techniques to estimate the incubation period can
be found in \cite{CCC89,BaG94}. The introduction of anti-retroviral
therapies and protease inhibitors during the 1990s in several parts
of the world resulted decline in opportunistic infections related
to AIDS \cite{Muga07,Conti2000,Hung2003}. As a result of such interventions,
the average incubation period was prolonged. There have been attempts
to estimate the incubation period that vary due to drug intervention
using statistical density functions \cite{ARZ92}. The impact of this
variation on the HIV dynamics, stability and on basic reproduction
number has been investigated \cite{CCC89,Kake89,CCC89-2}. In this
section, we first consider an ODE model that explains the dynamics
of HIV spread in a population leading to AIDS (see \cite{MaA88}).
We then consider a similar model where incubation period is a variable
with respect to a given therapy. We address issues of estimating incubation
period to be used in such dynamical models and the impact of above
mentioned therapies. Various ideas and the outline of this work are
given at the end of this section. 

Perhaps the most fundamental model for the epidemiology of AIDS is
that given by \cite{MaA88,RMA88,AAM91}, which takes the form

\begin{eqnarray}
\frac{dX}{dt} & = & \Lambda-\left(\lambda+\mu\right)X,\nonumber \\
\frac{dY}{dt} & = & \lambda X-\left(d+\mu\right)Y,\label{1.1}\\
\frac{dD_{z}}{dt} & = & dY-\gamma D_{z}.\nonumber 
\end{eqnarray}

Here the total population ($N$) is divided into susceptibles ($X$),
infectives ($Y$) and individuals with the full blown disease ($D_{z}$).
The parameter $\Lambda$ is the input into the susceptible class,
which can be defined as the number of births in the population, $\lambda$
is the force of infection, $\mu$ is general (non-AIDS related) mortality,
$\gamma$ is disease related mortality and $1/d$ is the average incubation
period. Here the incubation period is defined as the duration of time
between infection and onset of full blown disease. There are several
other constructions of HIV transmission dynamics models 

\begin{figure}
\includegraphics[scale=0.5]{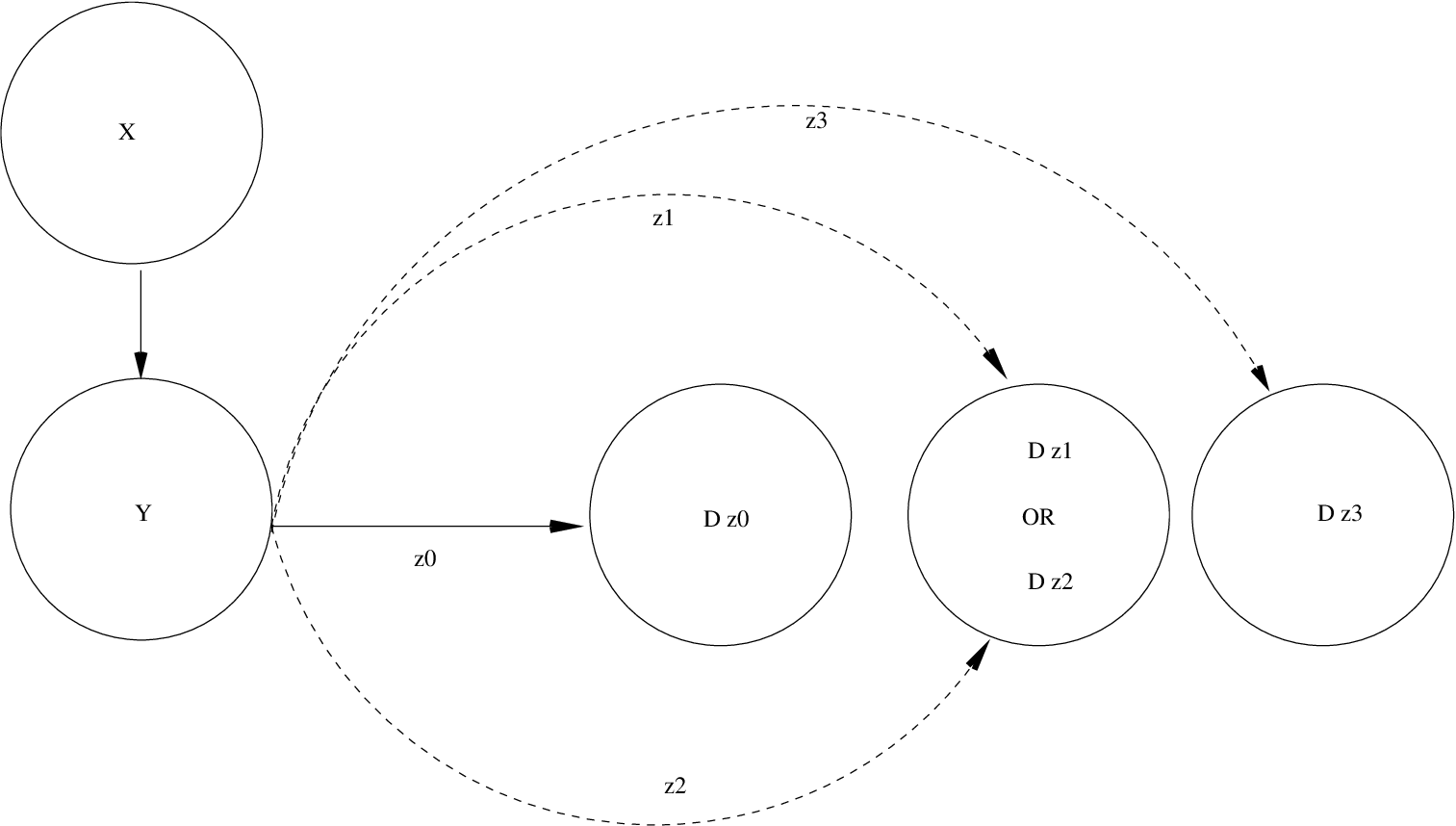}

\caption{Schematic diagram explaining the flow of infected individuals without
therapy to individuals who are on therapy.}
\label{fig-schematic-diagram}\label{}
\end{figure}

In the models involving the disease progression parameter, it has
been assumed that there is an increase in the mean length of life
after HIV infection since the availability of therapies for AIDS \cite{UN2006,Stov06}.
There are several works describing the impact of anti-retroviral therapies
using data \cite{AOO,mOU97,wEI95,OverM2006,Gray2011,Shaf2011,Reyn2011,Thiru2011,Mahy2010,Pretonis2010,Web2010,Eng09,Rajgop09,Sal2008,Johnson08,Hall08,Larson08,Boy07,Simc07,Muga07,Golub07}
and through models \cite{OverM2006,Stov06,Artz04,Rong07-1,Rong07-2}.
The time to start ART based on the CD4 count is still debatable. In
a recent study on HIV-1 discordant couples \cite{Cohen2011}, it was
observed that the incidence rates among early ART couples are lower
than than the incidence rates among couples who were given ART at
a standard time. Drugs are available which cannot eliminate virus
from the body, but are helpful in prolonging the life of an individual
by slowing the disease progression (in other words increasing the
incubation period). For example, protease inhibitors (say \emph{drug
1)} facilitate in producing non-infectious virus (only infectious
virus participates in new virus production), hence slowing the disease
progression; anti-retroviral therapy (\emph{say drug 2)} blocks virus
from interacting with the non-infected cells and hence reduces the
infection process within the cell population (see section 5 in \cite{Nowak2000}
and \cite{Perel99} for fuller details); and a combination of the
above two drugs (\emph{say drug 3}) can be more effective by simultaneously
combining the function of \emph{drug 1} and \emph{drug 2.} Note that,
when model (1.1) was developed, the above described drugs were not
available. Information on scale-up of anti-retroviral therapies and
related monitoring of individuals can be found elsewhere (for example,
see \cite{OverM2006,Boerma2006,Jogle2011,Srikant2010}. We assume
that once individuals start taking drugs, their average incubation
period is prolonged. So, instead of assuming a constant $1/d$, we
assume that it varies based on the drug type. Thus we define $1/d_{i}=\int_{\mathbb{R}}z_{i}g(z_{i})dz_{i},$
for $i=0,1,2,3,$ where $i=0$ denotes the without drug scenario,
$i=1$ for \emph{drug 1, $i=2$} for \emph{drug 2} and $i=3$ for
\emph{drug 3}. Here $g$ is the probability density function with
a certain parameter set (say $\bold B$) and $z_{i}$ is a continuous
random variable representing the incubation period. Here $z_{i}$
is a real valued function defined on a standard probability space
$\left(S,\mathbb{A},P\right)$, where $S$ is the space of elementary
events, $\mathbb{A}$ is called a Borel fields, and $P(\mathbb{A})$
is probability of the event $A\in\mathbb{A}.$ So, $z:S\rightarrow\mathbb{R}.$
We can also denote this integral as a Stieltjes integral $\int_{\mathbb{R}}z_{i}dG(z_{i})$,
where $G(z)=P(Z<z).$ We further assume without loss of generality
that

\begin{eqnarray}
\int_{\mathbb{R}}z_{0}dG(z_{0}) & < & \int_{\mathbb{R}}z_{1}dG(z_{1})\leq z_{2}dG(z_{2})<\int_{\mathbb{R}}z_{3}dG(z_{3}),\label{1.2}\\
\int_{\mathbb{R}}z_{0}dG(z_{0}) & < & \int_{\mathbb{R}}z_{1}dG(z_{1})>\int_{\mathbb{R}}z_{2}dG(z_{2})<\int_{\mathbb{R}}z_{3}dG(z_{3}).\nonumber 
\end{eqnarray}
(In the next section, we will give a detailed estimation procedure
for $\bold B.$) Applying these varying incubation periods, model
(\ref{1.1}) is modified as follows:

\begin{eqnarray}
\frac{dX}{dt} & = & \Lambda-\left(\lambda_{0}+\lambda_{1}+\lambda_{2}+\lambda_{3}+\mu\right)X,\nonumber \\
\frac{dY_{0}}{dt} & = & \lambda_{0}X-\left\{ \left(\int_{\mathbb{R}}z_{0}dG(z_{0})\right)^{-1}+\mu\right\} Y_{0},\nonumber \\
\frac{dY_{1}}{dt} & = & \lambda_{1}X-\left\{ \left(\int_{\mathbb{R}}z_{1}dG(z_{1})\right)^{-1}+\mu\right\} Y_{1},\nonumber \\
\frac{dY_{2}}{dt} & = & \lambda_{2}X-\left\{ \left(\int_{\mathbb{R}}z_{2}dG(z_{2})\right)^{-1}+\mu\right\} Y_{2},\nonumber \\
\frac{dY_{3}}{dt} & = & \lambda_{3}X-\left\{ \left(\int_{\mathbb{R}}z_{3}dG(z_{3})\right)^{-1}+\mu\right\} Y_{3},\nonumber \\
\frac{dD_{z_{0}}}{dt} & = & \left(\int_{\mathbb{R}}z_{0}dG(z_{0})\right)^{-1}Y_{0}-\gamma_{0}D_{z_{0}},\nonumber \\
\frac{dD_{z_{1}}}{dt} & = & \left(\int_{\mathbb{R}}z_{1}dG(z_{1})\right)^{-1}Y_{1}-\gamma_{1}D_{z_{1}},\label{1.3}\\
\frac{dD_{z_{2}}}{dt} & = & \left(\int_{\mathbb{R}}z_{2}dF(z_{2})\right)^{-1}Y_{2}-\gamma_{2}D_{z_{2}},\nonumber \\
\frac{dD_{z_{3}}}{dt} & = & \left(\int_{\mathbb{R}}z_{3}dG(z_{3})\right)^{-1}Y_{3}-\gamma_{3}D_{z_{3}},\nonumber 
\end{eqnarray}
where $Y_{0},Y_{1},Y{}_{2}$ and $Y_{3}$ are variables for infectives,
$D_{z_{0}},D_{z_{1}},D_{z_{2}}$ and $D_{z_{3}}$ are variables for
individuals with the full blown disease, $\lambda_{0},\lambda_{1},\lambda_{2}$
and $\lambda_{3}$and $\gamma_{0},\gamma_{1},\gamma_{2}$ and $\gamma_{3}$
are variables for disease related mortality for no-drug, \emph{drug
1, drug 2} and \emph{drug 3} respectively. See Figure \ref{fig-schematic-diagram}
describing the flows in the model eq (\ref{1.3}). General mortality
and disease-related mortality are incorporated in to the model to
demonstrate the basic structure of the model, and our aim here is
to estimate $\bold B$ and thus to estimate $\int_{\mathbb{R}}z_{i}dG(z_{i})$
for all $i$ such that simulations of the model are performed. In
model (\ref{1.3}), the total population $N=X+Y_{0}+Y_{1}+Y_{2}+Y_{3}+D_{z_{0}}+D_{z_{1}}+D_{z_{2}}+D_{z_{3}}$
satisfies 

\begin{eqnarray*}
\frac{dN}{dt} & = & \Lambda-\mu X+\mu\sum_{i=0}^{i=3}Y_{i}-\sum_{i=0}^{3}\gamma_{i}D_{z_{i}}.
\end{eqnarray*}

Estimation of parameters for the varying incubation periods is important
for understanding the impact of drugs in prolonging the onset of disease
and thus to prolong the life. The set $\bold B$ will also be useful
in obtaining varying basic reproductive rates, $R_{0i}$ for all $i=0,1,2,3$.
This can be computed as $R_{0i}=\lambda\gamma_{i}\int_{\mathbb{R}}z_{i}dG(z_{i})$
by assuming independence of the impact of various drugs. So far, there
is no evidence that $\beta,$ the probability of infecting a susceptible
partner changes with the activation of a drug in the body. If we assume
this as a constant, then $R_{00}\geq R_{01}\geq$ or $<R_{02}\geq R_{03}.$
$R_{01}<\left\{ R_{01},R_{02},R_{03}\right\} $, because individuals
are assumed to have longer incubation period due to the effect of
drugs. In the absence of clinical evidence, we assume that the impact
of \emph{drug1} and \emph{drug2} follows any one of the following
relations: $\int_{\mathbb{R}}z_{1}dG(z_{1})\leq$ or $>\textrm{ }\int_{\mathbb{R}}z_{2}dG(z_{2}).$
Similarly, another important epidemiological measure, the doubling
time, $t_{di}$ is obtained as $t_{di}=\ln(2)\int_{\mathbb{R}}z_{i}dG(z_{i})/\left[R_{0i}-1\right].$
Anti-retroviral therapy helps in blocking the virus from interacting
with cells and simultaneously providing protease inhibitors helps
in producing non-infectious virus. So without loss of generality,
it is assumed that the impact of double drug therapy is better than
a single drug therapy. If we assume disease related mortality is constant
for all $i$ then $\gamma_{i}=\gamma.$ The number of AIDS related
deaths in general are high during latter part of the incubation period
due to increase in opportunistic infections. In general, over all
AIDS related mortality rate in a population is assumed to be higher
than the general mortality rate in that population. Where there are
$n$ types of drugs available, we write the general form for the above
dynamical model as follows:

\begin{eqnarray}
\frac{dX}{dt} & = & \Lambda-\left(\sum_{I=0}^{n}\lambda_{i}+\mu\right)X,\nonumber \\
\frac{dY_{0}}{dt} & = & \lambda_{0}X-\left\{ \left(\int_{\mathbb{R}}z_{0}dG(z_{0})\right)^{-1}+\mu\right\} Y_{0},\nonumber \\
\vdots &  & \vdots\nonumber \\
\vdots &  & \vdots\nonumber \\
\frac{dY_{n}}{dt} & = & \lambda_{n}X-\left\{ \left(\int_{\mathbb{R}}z_{n}dG(z_{n})\right)^{-1}+\mu\right\} Y_{n},\nonumber \\
\frac{dD_{z_{0}}}{dt} & = & \left(\int_{\mathbb{R}}z_{0}dG(z_{0})\right)^{-1}Y_{0}-\gamma_{0}D_{z_{0}},\nonumber \\
\vdots &  & \vdots\label{1.4}\\
\vdots &  & \vdots\nonumber \\
\frac{dD_{z_{n}}}{dt} & = & \left(\int_{\mathbb{R}}z_{n}dG(z_{n})\right)^{-1}Y_{n}-\gamma_{n}D_{z_{n}},\nonumber 
\end{eqnarray}

As a special case we can consider all the parameters in the above
model as Stieltjes integrals and can estimate them using the rigorous
procedure explained in the next section. For $n=3$ in the model (\ref{1.4}),
we can deduce the model (\ref{1.3}). Practically, we do not have
a situation where several drugs are available in the market for HIV
infected individuals. Hence, the model (\ref{1.4}) should be treated
as a theoretical generalization. 

This paper is organized as follows: In section 2, we present contemporary
models constructed for understanding transmission dynamics of HIV
for policy formulations and corresponding modified disease progression
component that captures the impact of therapy, in section 3, we describe
in detail the estimation of the set $\bold B$ for up to three drugs;
section 4 gives the corresponding expressions for the conditional
probabilities of $N-$drugs. We construct theoretical examples using
three functions: Gamma, Logistic and Log-normal in section 5 to demonstrate
the method explained in section 3. In section 6 analysis for age-structured
populations is described in detail. Overall conclusions are given
in section 7. Appendix I gives equations for conditional probability
when incubation period for various drug types does not have the monotonicity
property. Appendix II gives some more theoretical examples when the
incubation period is truncated to the right. Appendix III provides
parameter values adopted for numerical simulations and Appendix IV
has numerical demonstration of the model outputs and sensitivity of
parameters in projecting HIV and AIDS.

\section{\textbf{Contemporary models and Modifications}}

In this section, we present a contemporary model for HIV epidemic
in India \cite{Rao2009} and corresponding integro-differential equations.
HIV model \cite{Rao2009} developed based on Indian data has three
components, 1) Model for spread in general population, 2) model for
spread in homosexual men (MSM), 3) model for spread in intravenous
drug users (IDU). We provide a description of this model and then
write corresponding revised model with integro differential equations.
The system of differential equations in three models have incorporated
dynamics in fourteen compartments: $U(i),$ susceptible population;
$V(i),$ sexually transmitted diseases population; $W(i),$ HIV infected;
$T(i),$ AIDS in the general population for gender $i$ (say, $i=1$
for male and $i=2$ for female), $U(m),$ susceptible MSM; $V(m),$
sexually transmitted infected MSM;, $W(m),$ HIV infected MSM; $T(m),$
MSM population with AIDS; $U(IDU),$ susceptible intravenous drug
users; $W(IDU),$ HIV infected intravenous drug users; $T(IDU),$
intravenous drug users with AIDS. Male susceptible in general population
is eligible to acquire virus from $j$ th sub-population ($j=1,$
female married partner; $j=2,$ female casual partner, $j=3,$ commercial
sex worker; $j=4,$ through blood transfusions. All the sub-populations
are allowed to contribute for the transmission dynamics of HIV and
each sub-population is also subject to the risk of acquiring the infection
from other sub-population wherever applicable (see \cite{Rao2009}
for complete description). 

The differential equations describing the Indian HIV epidemic model
are

\begin{eqnarray*}
\begin{array}{c}
\mbox{General}\\
\mbox{Model}
\end{array} & \begin{cases}
\frac{dU(i)}{dt} & =a_{i}U(i)-U(i)\left(\sum_{j=1}^{4}\frac{b_{ij}V(i)}{N(j)}+\sum_{j=1}^{4}\frac{c_{ij}W(j)}{N(j)}\right)+\phi V(i)\\
\\
\frac{dV(i)}{dt} & =U(i)\sum_{j=1}^{4}\frac{b_{ij}V(i)}{N(j)}-V(i)\sum_{j=1}^{4}\frac{d_{ij}W(j)}{N(j)}-\mu V(i)-\phi V(i)\\
\\
\frac{dW(i)}{dt} & =U(i)\sum_{j=1}^{4}\frac{c_{ij}W(j)}{N(j)}+V(i)\sum_{j=1}^{4}\frac{d_{ij}W(j)}{N(j)}-\\
\\
 & \qquad\qquad\qquad\qquad\qquad\qquad\qquad\delta_{i}W(i)-\alpha iW(i)\\
\\
\frac{dT(i)}{dt} & =\alpha_{i}W(i)-\mu_{i}T(i)
\end{cases}
\end{eqnarray*}

\begin{eqnarray*}
\begin{array}{c}
\mbox{MSM}\\
\mbox{Model}
\end{array} & \begin{cases}
\frac{dU(m)}{dt} & =a_{m}U(m)-U(m)\left(\frac{b_{m}V(m)}{N(m)}+\frac{c_{m}W(m)}{N(m)}\right)+\phi V(m)\\
\\
\frac{dV(m)}{dt} & =U(m)\frac{b_{m}V(m)}{N(m)}-V(m)\frac{d_{m}W(m)}{N(m)}-\mu V(m)-\phi V(m)\\
\\
\frac{dW(m)}{dt} & =U(m)\frac{c_{m}W(m)}{N(m)}+V(m)\frac{d_{m}W(m)}{N(m)}-\\
\\
 & \qquad\qquad\qquad\qquad\qquad\qquad\delta_{m}W(m)-\alpha_{m}W(m)\\
\\
\frac{dT(m)}{dt} & =\alpha_{m}W(m)-\mu_{m}T(m)
\end{cases}
\end{eqnarray*}

\begin{eqnarray*}
\begin{array}{c}
\mbox{IDU}\\
\mbox{Model}
\end{array} & \begin{cases}
\frac{dU(IDU)}{dt} & =a_{IDU}U(IDU)-U(IDU)\left(\frac{c_{IDU}W(IDU)}{N(IDU)}\right)+\\
\\
 & \qquad\qquad\qquad\qquad\qquad\qquad\qquad\phi V(IDU)\\
\\
\frac{dW(IDU)}{dt} & =U(IDU)\frac{c_{IDU}W(IDU)}{N(IDU)}-\delta_{IDU}W(IDU)-\\
\\
 & \qquad\qquad\qquad\qquad\qquad\qquad\qquad\alpha_{IDU}W(IDU)\\
\\
\frac{dT(IDU)}{dt} & =\alpha_{IDU}T(IDU)-\mu_{IDU}T(IDU)
\end{cases}
\end{eqnarray*}

where, $N(j)=V(j)+W(j)$ for $j=1,2,3,4$ and $N(m)=V(m)+W(m)$

The corresponding models with flexible incubation periods are

\begin{eqnarray*}
\frac{dU(i)}{dt} & = & a_{i}U(i)-U(i)\left(\sum_{j=1}^{4}\frac{b_{ij}V(i)}{N(j)}+\sum_{k=0}^{3}\sum_{j=1}^{4}\frac{c_{ijk}W(j)}{N(j)}\right)+\phi V(i)\\
\frac{dV(i)}{dt} & = & U(i)\sum_{j=1}^{4}\frac{b_{ij}V(i)}{N(j)}-V(i)\sum_{k=0}^{3}\sum_{j=1}^{4}\frac{d_{ijk}W(j)}{N(j)}-\mu V(i)-\phi V(i)\\
\frac{dW_{0}(i)}{dt} & = & U(i)\sum_{j=1}^{4}\frac{c_{ij0}W(j)}{N(j)}+V(i)\sum_{j=1}^{4}\frac{d_{ij0}W(j)}{N(j)}-\\
 &  & \qquad\qquad\qquad\qquad\qquad\left\{ \left(\int_{\mathbb{R}}z_{0}dG(z_{0})\right)^{-1}+\delta_{i}\right\} W_{0}(i)\\
\frac{dW_{1}(i)}{dt} & = & U(i)\sum_{j=1}^{4}\frac{c_{ij1}W(j)}{N(j)}+V(i)\sum_{j=1}^{4}\frac{d_{ij1}W(j)}{N(j)}-\\
 &  & \qquad\qquad\qquad\qquad\qquad\left\{ \left(\int_{\mathbb{R}}z_{1}dG(z_{1})\right)^{-1}+\delta_{i}\right\} W_{1}(i)\\
\frac{dW_{2}(i)}{dt} & = & U(i)\sum_{j=1}^{4}\frac{c_{ij2}W(j)}{N(j)}+V(i)\sum_{j=1}^{4}\frac{d_{ij2}W(j)}{N(j)}-\\
 &  & \qquad\qquad\qquad\qquad\qquad\left\{ \left(\int_{\mathbb{R}}z_{2}dG(z_{2})\right)^{-1}+\delta_{i}\right\} W_{2}(i)\\
\frac{dW_{3}(i)}{dt} & = & U(i)\sum_{j=1}^{4}\frac{c_{ij3}W(j)}{N(j)}+V(i)\sum_{j=1}^{4}\frac{d_{ij3}W(j)}{N(j)}-\\
 &  & \qquad\qquad\qquad\qquad\qquad\left\{ \left(\int_{\mathbb{R}}z_{3}dG(z_{3})\right)^{-1}+\delta_{i}\right\} W_{3}(i)\\
\frac{dT_{z0}(i)}{dt} & = & \left(\int_{\mathbb{R}}z_{0}dG(z_{0})\right)^{-1}W_{0}(i)-\mu_{0}T_{zo}\\
\frac{dT_{z1}(i)}{dt} & = & \left(\int_{\mathbb{R}}z_{1}dG(z_{1})\right)^{-1}W_{1}(i)-\mu_{1}T_{z1}\\
\frac{dT_{z2}(i)}{dt} & = & \left(\int_{\mathbb{R}}z_{2}dG(z_{2})\right)^{-1}W_{2}(i)-\mu_{1}T_{z1}\\
\frac{dT_{z3}(i)}{dt} & = & \left(\int_{\mathbb{R}}z_{3}dG(z_{3})\right)^{-1}W_{3}(i)-\mu_{3}T_{z3}
\end{eqnarray*}

\begin{eqnarray*}
\frac{dU(m)}{dt} & = & a_{m}U(m)-U(m)\left(\frac{b_{m}V(m)}{N(m)}+\sum_{k=0}^{3}\frac{c_{m}W(m)}{N(m)}\right)+\phi V(m)\\
\frac{dV(m)}{dt} & = & U(m)\frac{b_{m}V(m)}{N(m)}-V(m)\sum_{k=0}^{3}\frac{d_{m}W(m)}{N(m)}-\mu V(m)-\phi V(m)\\
\frac{dW_{0}(m)}{dt} & = & U(m)\frac{c_{m0}W(m)}{N(m)}+V(m)\frac{d_{m0}W(m)}{N(m)}-\\
 &  & \qquad\qquad\qquad\qquad\qquad\left\{ \left(\int_{\mathbb{R}}z_{0}dG(z_{0})\right)^{-1}+\delta_{m}\right\} W_{0}(m)\\
\frac{dW_{1}(m)}{dt} & = & U(m)\frac{c_{m1}W(m)}{N(m)}+V(m)\frac{d_{m1}W(m)}{N(m)}-\\
 &  & \qquad\qquad\qquad\qquad\qquad\left\{ \left(\int_{\mathbb{R}}z_{1}dG(z_{1})\right)^{-1}+\delta_{m}\right\} W_{1}(m)\\
\frac{dW_{2}(m)}{dt} & = & U(m)\frac{c_{m2}W(m)}{N(m)}+V(m)\frac{d_{m2}W(m)}{N(m)}-\\
 &  & \qquad\qquad\qquad\qquad\qquad\left\{ \left(\int_{\mathbb{R}}z_{2}dG(z_{2})\right)^{-1}+\delta_{m}\right\} W_{2}(m)\\
\frac{dW_{3}(m)}{dt} & = & U(m)\frac{c_{m3}W(m)}{N(m)}+V(m)\frac{d_{m3}W(m)}{N(m)}-\\
 &  & \qquad\qquad\qquad\qquad\qquad\left\{ \left(\int_{\mathbb{R}}z_{0}dG(z_{0})\right)^{-1}+\delta_{m}\right\} W_{3}(m)\\
\frac{dT_{z0}(m)}{dt} & = & \left(\int_{\mathbb{R}}z_{0}dG(z_{0})\right)^{-1}W_{0}(m)-\mu_{m}T_{z0}(m)\\
\frac{dT_{z1}(m)}{dt} & = & \left(\int_{\mathbb{R}}z_{1}dG(z_{1})\right)^{-1}W_{1}(m)-\mu_{m}T_{z1}(m)\\
\frac{dT_{z2}(m)}{dt} & = & \left(\int_{\mathbb{R}}z_{2}dG(z_{2})\right)^{-1}W_{2}(m)-\mu_{m}T_{z2}(m)\\
\frac{dT_{z3}(m)}{dt} & = & \left(\int_{\mathbb{R}}z_{3}dG(z_{3})\right)^{-1}W_{3}(m)-\mu_{m}T_{z3}(m)
\end{eqnarray*}

\begin{eqnarray*}
\frac{dU(IDU)}{dt} & = & a_{IDU}U(IDU)-U(IDU)\sum_{k=0}^{3}\left(\frac{c_{IDU,k}W(IDU)}{N(IDU)}\right)+\phi V(IDU)\\
\frac{dW_{0}(IDU)}{dt} & = & U(IDU)\left(\frac{c_{IDU,0}W(IDU)}{N(IDU)}\right)-\\
 &  & \qquad\qquad\qquad\qquad\qquad\left\{ \left(\int_{\mathbb{R}}z_{0}dG(z_{0})\right)^{-1}+\delta_{IDU}\right\} W_{0}(IDU)\\
\frac{dW_{1}(IDU)}{dt} & = & U(IDU)\left(\frac{c_{IDU,1}W(IDU)}{N(IDU)}\right)-\\
 &  & \qquad\qquad\qquad\qquad\qquad\left\{ \left(\int_{\mathbb{R}}z_{1}dG(z_{1})\right)^{-1}+\delta_{IDU}\right\} W_{1}(IDU)\\
\frac{dW_{2}(IDU)}{dt} & = & U(IDU)\left(\frac{c_{IDU,2}W(IDU)}{N(IDU)}\right)-\\
 &  & \qquad\qquad\qquad\qquad\qquad\left\{ \left(\int_{\mathbb{R}}z_{2}dG(z_{2})\right)^{-1}+\delta_{IDU}\right\} W_{2}(IDU)\\
\frac{dW_{3}(IDU)}{dt} & = & U(IDU)\left(\frac{c_{IDU,3}W(IDU)}{N(IDU)}\right)-\\
 &  & \qquad\qquad\qquad\qquad\qquad\left\{ \left(\int_{\mathbb{R}}z_{3}dG(z_{3})\right)^{-1}+\delta_{IDU}\right\} W_{3}(IDU)\\
\frac{dT_{z0}(IDU)}{dt} & = & \left(\int_{\mathbb{R}}z_{0}dG(z_{0})\right)^{-1}W_{0}(IDU)-\mu_{IDU}T_{z0}(IDU)\\
\frac{dT_{z1}(IDU)}{dt} & = & \left(\int_{\mathbb{R}}z_{1}dG(z_{1})\right)^{-1}W_{1}(IDU)-\mu_{IDU}T_{z1}(IDU)\\
\frac{dT_{z2}(IDU)}{dt} & = & \left(\int_{\mathbb{R}}z_{2}dG(z_{2})\right)^{-1}W_{2}(IDU)-\mu_{IDU}T_{z2}(IDU)\\
\frac{dT_{z3}(IDU)}{dt} & = & \left(\int_{\mathbb{R}}z_{3}dG(z_{3})\right)^{-1}W_{3}(IDU)-\mu_{IDU}T_{z3}(IDU)
\end{eqnarray*}

where the variables with suffix $z_{0},$ $z_{2}$, $z_{3}$, $z_{3}$
are corresponding to the impact drug0, drug1, drug2, drug3 respectively.
These contemporary models are improvised version of basic models presented
in section 1 and are tested to predict accurately the epidemic situation
during the era of anti-retroviral therapies.

\section{\textbf{Conditional probabilities}}

In this section, we will give a detailed procedure to estimate $\bold B$
through a deconvolution technique. Let $\bold B$ be split into a
collection of four parameter sets say, $\bold B$=$\left\{ B_{0},B_{1},B_{2},B_{3}\right\} $
for the four types of scenarios described in the previous section.
Let $H$ be the time of infection and $Z$ be the incubation period,
then the time of onset of the disease can be represented as $D=H+Z.$
There have been studies (see for list of references Brookmeyer and
Gail (1994)), in which $H$ and $Z$ were assumed independent and
$D$ was estimated through convolution. We outline the general idea
of convolution and then give the convolution of $H$ and $Z.$ Suppose
$(a_{n})$ and $(b_{n})$ are two sequences of numbers over the time
period, then

\begin{eqnarray}
(a_{n})*(b_{n}) & = & \sum_{k=0}^{n}a_{n}b_{n-k}\label{2.1}
\end{eqnarray}
where $(a_{n})*(b_{n})$ is the convolution of these sequences with
an operator $'*'$. Suppose $a$ and $b$ are mutually independent
random variables and let $A_{\mathcal{L}}(x)$ and $B_{\mathcal{L}}(x)$
be their Laplace transformations, then $a+b$ has the Laplace transformation
$A_{\mathcal{L}}B_{\mathcal{L}}.$ Since the multiplication of the
Laplace transformation is associative and commutative, it follows
that $(a_{n})*(b_{n})$ is also associative and commutative. Instead
of discrete notation, suppose $a$ and $b$ are continuous and independent
with probability density functions $h$ and $g$, then the density
of $h+g$ is given by 

\begin{eqnarray*}
f(s) & = & \int_{-\infty}^{\infty}h(t-s)g(t)dt=\int_{-\infty}^{\infty}h(t)g(t-s)ds
\end{eqnarray*}
.

Suppose $G(s)=\int_{-\infty}^{s}g(s)ds$, and $F(s)=\int_{-\infty}^{s}f(s)ds$
then 

\begin{eqnarray}
F(s) & = & \int_{-\infty}^{\infty}h(t)G(t-s)ds.\label{2.2}
\end{eqnarray}

We call $F$ the convolution of $h$ and $G$. Suppose the above $h$
and $G$ represent the infection density and incubation period distribution
function; then the convolution of $h$ and $G$ represents the cumulative
number of disease cases reported (or observed), and is given by 

\begin{eqnarray}
h*G & = & \int_{-\infty}^{\infty}h(t)G(t-s)ds.\label{2.3}
\end{eqnarray}

This kind of convolution in (\ref{2.3}) was used to estimate the
number of AIDS cases for the first time by \cite{BaG88}. Information
on $G$ may not be available for some populations. In such situations,
$G$ has been estimated through deconvolution from the information
available on $h*G$ and $h$ \cite{RaoASRS2004,RaoASRS2005-1}. In
this section we will construct conditional probabilities for each
\emph{drug type} and express the function that maximizes $\bold B$.
These kind of conditional probabilities derived for the \emph{drug}
\emph{type} were not available earlier for the incubation periods
when the total number of reported disease cases were considered. Note
that $h*G$ is the cumulative number of disease cases. 

Let $X_{0},X_{1},X_{2},...,X_{n-k},...,X_{n-l},...,X_{n-m},...,X_{n}$
be the disease cases available in the time intervals $[U_{i-1},U_{i})$
for $i=0,1,2,...n-k,...n-l,...(n-m)+1,...n+1.$ Suppose $\bold E$
is the event of diagnosis of disease after the first infection at
$T_{0}.$ Let $\bold E=\left\{ E_{0},E_{1},E_{2},E_{3}\right\} $
and $E_{0}$ occurs in the interval $[U_{0},U_{n-k})$, $E_{1}$ (or
$E_{2}$) in $[U_{n-k},U_{n-l})$ (or $[U_{n-l},U_{n-m})$) and $E_{3}$
in $[U_{n-m},U_{n}).$ Now $D,$ the cumulative number of disease
cases up to time $U_{n},$ can be expressed from (\ref{2.3}) as follows:

\begin{eqnarray}
D\left(U_{0}\leq s\leq U_{n}\right) & = & \int_{0}^{U_{n-k}}h(t)G(t-s)ds+\int_{U_{n-k}}^{U_{n-l}}h(t)G(t-s)ds\nonumber \\
\nonumber \\
 &  & +\int_{U_{n-l}}^{U_{n-m}}h(t)G(t-s)ds+\int_{U_{n-m}}^{U_{n}}h(t)G(t-s)ds,\label{2.4}
\end{eqnarray}
\\
$D\left(\bold A,\bold B/U_{n}\right)=\int_{0}^{U_{n-k}}h(t/A_{0})G(t-s/B_{0})ds+\int_{U_{n-k}}^{U_{n-l}}h(t/A_{1})G(t-s/B_{1})ds$ 

$ $

$\qquad\qquad+\int_{U_{n-l}}^{U_{n-m}}h(t/A_{2})G(t-s/B_{2})ds+\int_{U_{n-m}}^{U_{n}}h(t/A_{3})G(t-s/B_{3})ds$.

$ $

In the above equation $A_{0},$$A_{1},$ $A_{2}$ and $A_{3}$ are
the parameter sets for the $h$ for $drug0$, $drug1$, $drug2$ and
$drug3$. An infected individual could fall in to any of the intervals
described above, and similarly a full-blown disease diagnosed individual
could fall in the same interval, but for a given individual the chronological
time of infection would be earlier than that of diagnosis of the disease.
$U_{n-k}$ is the time of introduction of \emph{drugs} after infection
at $U_{0}.$ Individuals who were diagnosed on or after $U_{n-k},$
and before $U_{n},$ were taking one of the three \emph{drugs.} If
$E_{1}\in[U_{n-k},U_{n-l})$ and $E_{2}\in[U_{n-l},U_{n-m})$ $Z_{1}<Z_{2}$,
otherwise if $E_{1}\in[U_{n-l},U_{n-m})$ and $E_{2}\in[U_{n-k},U_{n-l})$
then and if $E_{1},E_{2}\in[U_{n-k},U_{n-m})$ then $Z_{1}=Z_{2}.$
An individual who was diagnosed with the disease before $U_{n}$ must
have developed symptoms in one of the four intervals $[U_{0},U_{n-k})$,
$[U_{n-k},U_{n-l})$, $[U_{n-l},U_{n-m})$ and $[U_{n-m},U_{n}).$
Let $E_{j}\in[U_{j-1},U_{j})\subseteq[U_{0},U_{n-k})$, then the conditional
probability of the occurrence of $E_{j}$ given $E$ is expressed
as

\begin{eqnarray*}
P\left(E_{j}/E\right) & = & P\left(U_{j-1}\leq D\leq U_{j}/D\leq U_{n}\right)\\
\\
 &  & =\frac{D\left(A_{0},B_{0}/U_{j}\right)-D\left(A_{0},B_{0}/U_{j-1}\right)}{D\left(A_{0},B_{0}/U_{n}\right)}
\end{eqnarray*}

\begin{eqnarray}
 & = & \int_{0}^{U_{j}}h(t/A_{0})G(t-s/B_{0})ds.\left[\int_{0}^{U_{n}}h(t/A_{0})G(t-s/B_{0})ds\right]^{-1}\nonumber \\
\nonumber \\
 &  & -\int_{0}^{U_{j-1}}h(t/A_{0})G(t-s/B_{0})ds.\left[\int_{0}^{U_{n}}h(t/A_{0})G(t-s/B_{0})ds\right]^{-1}.\label{2.5}
\end{eqnarray}

If \emph{drugs} were initiated at $U_{n-k},$ then these conditional
probabilities constructed above will change according to the occurrence
of $E_{1},E_{2},E_{3}.$ Consider $E_{1}\cap E_{2}=\emptyset.$ Let
$E_{k}\in[U_{k-1},U_{k})\subseteq[U_{n-k},U_{n-l})$, and $E_{1}\in[U_{n-k},U_{n-l})$,
then 

\begin{eqnarray*}
P\left(E_{k}/E\right) & = & P\left(U_{k-1}\leq D\leq U_{k}/D\leq U_{n}\right)\\
\\
 &  & =\frac{D\left(A_{1},B_{1}/U_{k}\right)-D\left(A_{1},B_{1}/U_{k-1}\right)}{D\left(A_{1},B_{1}/U_{n}\right)}
\end{eqnarray*}

\begin{eqnarray}
 & = & \int_{0}^{U_{k}}h(t/A_{1})G(t-s/B_{1})ds.\left[\int_{0}^{U_{n}}h(t/A_{1})G(t-s/B_{1})ds\right]^{-1}\nonumber \\
\nonumber \\
 &  & -\int_{0}^{U_{k-1}}h(t/A_{1})G(t-s/B_{1})ds.\left[\int_{0}^{U_{n}}h(t/A_{1})G(t-s/B_{1})ds\right].^{-1}\label{2.6}
\end{eqnarray}

Suppose $E_{k}\in[U_{k-1},U_{k})\subseteq[U_{n-k},U_{n-l})$, and
$E_{2}\in[U_{n-k},U_{n-l})$ i.e a situation when $Z_{1}>Z_{2}$,
then 

\begin{eqnarray}
P\left(E_{k}/E\right) & = & \int_{0}^{U_{k}}h(t/A_{2})G(t-s/B_{2})ds.\left[\int_{0}^{U_{n}}h(t/A_{2})G(t-s/B_{2})ds\right]^{-1}\nonumber \\
\nonumber \\
 &  & -\int_{0}^{U_{k-1}}h(t/A_{2})G(t-s/B_{2})ds.\left[\int_{0}^{U_{n}}h(t/A_{2})G(t-s/B_{2})ds\right]^{-1}.\label{2.7}
\end{eqnarray}

Let $E_{l}\in[U_{l-1},U_{l})\subseteq[U_{n-l},U_{n-m})$, and $E_{2}\in[U_{n-l},U_{n-m}),$
then

\begin{eqnarray*}
P\left(E_{l}/E\right) & = & P\left(U_{l-1}\leq D\leq U_{l}/D\leq U_{n}\right)\\
\\
 &  & =\frac{D\left(A_{2},B_{2}/U_{l}\right)-D\left(A_{2},B_{2}/U_{l-1}\right)}{D\left(A_{2},B_{2}/U_{n}\right)}
\end{eqnarray*}

\begin{eqnarray}
 & = & \int_{0}^{U_{l}}h(t/A_{2})G(t-s/B_{2})ds.\left[\int_{0}^{U_{n}}h(t/A_{2})G(t-s/B_{2})ds\right]^{-1}\nonumber \\
\nonumber \\
 &  & -\int_{0}^{U_{l-1}}h(t/A_{2})G(t-s/B_{2})ds.\left[\int_{0}^{U_{n}}h(t/A_{2})G(t-s/B_{2})ds\right].^{-1}\label{2.8}
\end{eqnarray}

Suppose $E_{l}\in[U_{l-1},U_{l})\subseteq[U_{n-l},U_{n-m})$, and
$E_{1}\in[U_{n-l},U_{n-m})$ i.e a situation when $Z_{1}>Z_{2},$
then 

\begin{eqnarray}
P\left(E_{l}/E\right) & = & \int_{0}^{U_{l}}h(t/A_{1})G(t-s/B_{1})ds.\left[\int_{0}^{U_{n}}h(t/A_{1})G(t-s/B_{1})ds\right]^{-1}\nonumber \\
\nonumber \\
 &  & -\int_{0}^{U_{l-1}}h(t/A_{1})G(t-s/B_{1})ds.\left[\int_{0}^{U_{n}}h(t/A_{1})G(t-s/B_{1})ds\right]^{-1}.\label{2.9}
\end{eqnarray}

Now consider $E_{1}=E_{2}\in[U_{p-1},U_{p})\subseteq[U_{n-k},U_{n-m}),$
i.e. $Z_{1}=Z_{2},$ then the conditional probabilities contain the
same parameter sets. In this situation,

\begin{eqnarray}
P\left(E_{p}/E\right) & = & \int_{0}^{U_{p}}h(t/A_{1})G(t-s/B_{1})ds.\left[\int_{0}^{U_{n}}h(t/A_{1})G(t-s/B_{1})ds\right]^{-1}\nonumber \\
\nonumber \\
 &  & -\int_{0}^{U_{p-1}}h(t/A_{1})G(t-s/B_{1})ds.\left[\int_{0}^{U_{n}}h(t/A_{1})G(t-s/B_{1})ds\right]^{-1}.\label{2.10}
\end{eqnarray}

Since $Z_{3}>Z_{0},Z_{1},Z_{2}$, suppose $E_{3}\in[U_{m-1},U_{m})\subseteq[U_{n-m},U_{n}]$,
then 

\begin{eqnarray*}
 &  & P\left(U_{m-1}\leq D\leq U_{m}/D\leq U_{n}\right)=\frac{D\left(A_{3},B_{3}/U_{m}\right)-D\left(A_{3},B_{3}/U_{m-1}\right)}{D\left(A_{3},B_{3}/U_{n}\right)}.
\end{eqnarray*}
Therefore,

\begin{eqnarray}
P\left(E_{m}/E\right) & = & \int_{0}^{U_{m}}h(t/A_{3})G(t-s/B_{3})ds.\left[\int_{0}^{U_{n}}h(t/A_{3})G(t-s/B_{3})ds\right]^{-1}\nonumber \\
\nonumber \\
 &  & -\int_{0}^{U_{m-1}}h(t/A_{3})G(t-s/B_{3})ds.\left[\int_{0}^{U_{n}}h(t/A_{3})G(t-s/B_{3})ds\right]^{-1}\label{2.11}
\end{eqnarray}
The above conditional probabilities $P\left(E_{j}/E\right),P\left(E_{k}/E\right),$
$P\left(E_{l}/E\right),$ $P\left(E_{p}/E\right)$ and $P\left(E_{m}/\bold E\right)$
are the probabilities associated with the intervals $[U_{j'-1},U_{j'})$
$,[U_{k'-1},U_{k'}),$ $[U_{l'-1},U_{l'}),[U_{p'-1},U_{p'})$ and
$[U_{m'-1},U_{m'})$ for the ranges of $j,k,l,p$ and $m$ defined
above. Since, $X_{0},X_{1},X_{2},...,X_{n-k},...,$$X_{n-l},...,X_{n-m},...,X_{n}$
are mutually exclusive, we assume they follow a parametric distribution
with the above probabilities are mutually exclusive, so we assume
they follow the multinomial property of the distribution of the values
in the time intervals and the above conditional probabilities. Then
the likelihood functions corresponding to the event set $\bold E$
are $L_{0}\left(\bold A,\bold B/P_{j}\right)=$$\prod_{j'=1}^{n-k}$
$P_{j}\left(\bold A,\bold B/T_{j'}\right)$, $L_{1(2)}\left(\bold A,\bold B/P_{k}\right)=\prod_{k'=n-k}^{n-l}$$P_{k'}\left(\bold A,\bold B/T_{k'}\right)$,
$L_{2(1)}\left(\bold A,\bold B/P_{l'}\right)=\prod_{l'=n-l}^{n-m}$
$P_{l'}\left(\bold A,\bold B/T_{l'}\right)$, $L_{1=2}\left(\bold A,\bold B/P_{p}\right)$
$=\prod_{p'=n-k}^{n-m}P_{p'}\left(\bold A,\bold B/T_{p'}\right)$
and $ $ $L_{3}\left(\bold A,\bold B/P_{m}\right)=\prod_{m'=n-m}^{n}P_{m'}\left(\bold A,\bold B/T_{m'}\right).$
Here $P_{\bullet}=P(E_{\bullet}/\bold E).$ We estimate $\bold A$
by fitting an infection curve from the incidence data and we then
estimate $\bold B$ by maximizing the likelihood functions expressed
above. The best estimate of $\bold A$ could be information for initial
values of $X$ and $Y$ in the model (\ref{1.3}). Using the corresponding
estimate of $\bold B$, we obtain $\int_{\mathbb{R}}z_{i}dF(z_{i}).$
In such situations, the above likelihood functions would be $L_{0}=\prod_{j'=0}^{n-k}P_{j}^{T_{j'}},$
$L_{1(2)}=\prod_{k'=n-k}^{n-l}P_{k}^{T_{k'}},$$L_{2(1)}=\prod_{l'=n-l}^{n-m}P_{l}^{T_{l'}},$
$L_{1=2}=\prod_{p'=n-k}^{n-m}P_{p}^{T_{p'}}$ and $L_{3}=$ $\prod_{m'=n-m}^{n}P_{m}^{T_{m'}}.$

\section{\textbf{Generalization for multiple drug impact}}

In this section, expressions for the conditional probabilities are
presented when multiple drugs are administrated in the population.
Refer to the sections 2 and 3 for introduction on the role of various
drugs and refer to the section 4 for basic formulations of conditional
probabilities when there are three types of drugs to prolong the incubation
period and without any drug situation that would not alter natural
process of disease progression. Modeling for the situation corresponding
to no drug is highly relevant for those countries where surveillance
and diagnosis of infections are not complete and several individuals
with HIV are not taking drugs. Let $N=\left\{ N_{0},N_{1},N_{2},\right.$
$\left....,N_{N}\right\} $ be the number of available drugs and $Z=\left\{ Z_{0},Z_{1},Z_{2},...,Z_{N}\right\} $
be their corresponding incubation periods. Further let $Z_{0}<Z_{1}<Z_{2}<,...,<Z_{N}$
and $\bold A$ and $\bold B$ be their parametric sets. Then 

\begin{eqnarray}
D\left(\bold A,\bold B/U_{N_{N}}\right) & = & \int_{0}^{U_{N_{0}}}h(t/A_{N_{0}})G(t-s/B_{N_{0}})ds\nonumber \\
\nonumber \\
 &  & +\int_{U_{N_{0}}}^{U_{N_{1}}}h(t/A_{N_{1}})G(t-s/B_{N_{1}})ds\nonumber \\
\nonumber \\
 &  & \cdots+\int_{U_{N_{N-1}}}^{U_{N_{N}}}h(t/A_{N_{N}})G(t-s/B_{N_{N}})ds.\label{three-one}
\end{eqnarray}

Now, $P\left(E_{N_{i}}/E\right)=P\left(U_{N_{i-1}}\leq D\leq U_{N_{i}}/D\leq U_{N_{N}}\right)$
and $L_{N_{i}}$ (for some $i$) can be computed as follows:

\begin{eqnarray}
P\left(E_{N_{i}}/E\right) & = & \int_{0}^{U_{N_{i}}}h(t/A_{N_{i}})G(t-s/B_{N_{i}})ds\times\nonumber \\
\nonumber \\
 &  & \left[\int_{0}^{U_{N_{N}}}h(t/A_{N_{N}})G(t-s/B_{N_{N}})ds\right]^{-1}\nonumber \\
\nonumber \\
 &  & -\int_{0}^{U_{N_{i-1}}}h(t/A_{N_{i-1}})G(t-s/B_{N_{i-1}})ds\times\nonumber \\
\nonumber \\
 &  & \left[\int_{0}^{U_{N_{N}}}h(t/A_{N_{N}})G(t-s/B_{N_{N}})ds\right].^{-1}\label{three-two}
\end{eqnarray}

$L_{N_{i}}=\prod_{j=N_{i-1}}^{N_{i}}P_{j}^{T_{j}}$ is maximized for
the set $\left[A_{i},B_{i}\right]$ by the procedure explained in
the previous section. We will obtain $N$ sets of $\left[\bold A,\bold B\right]$
values, and the corresponding likelihood values are $L_{N_{1}},L_{N_{2}},L_{N_{3}},...,L_{N_{N}}.$
In the above, we have assumed monotonicity of $(Z_{i})$ to arrive
at (\ref{three-two}). If the $(Z_{i})$ values are not monotonic
then the various conditional probabilities can be constructed as explained
in the previous section. There we explained the general expression
when there were a finite number of drugs available on the market.
A detailed construction of various conditional probabilities is not
necessary for the purpose of the present section (for details see
Appendix I). When the $Z_{i}s$ are not monotonic, and if they follow
some order, say for example, $Z_{0}>Z_{1}<Z_{2}>...<Z_{N},$ then
the conditional probabilities can be constructed in the same way as
equations (\ref{2.7}e\ref{2.9}) were. Suppose $(Z_{p})$ are equal
for each $p$, then there will be two scenarios arising: one for before
drug intervention and one after drug intervention. For this situation,
the likelihood equation is $L_{N_{p}}=\prod_{p=N_{p-1}}^{N_{p}}P_{p}^{T_{p}}$
where $P\left(E_{N_{p}}/E\right),{}$ is given as follows:

\begin{eqnarray}
P\left(E_{N_{p}}/E\right) & = & \int_{0}^{U_{N_{p}}}h(t/A_{N_{p}})G(t-s/B_{N_{p}})ds\times\nonumber \\
\nonumber \\
 &  & \left[\int_{0}^{U_{N_{N}}}h(t/A_{N_{N}})G(t-s/B_{N_{N}})ds\right]^{-1}\nonumber \\
\nonumber \\
 &  & -\int_{0}^{U_{N_{p-1}}}h(t/A_{N_{p-1}})G(t-s/B_{N_{p-1}})ds\times\nonumber \\
\nonumber \\
 &  & \left[\int_{0}^{U_{N_{N}}}h(t/A_{N_{N}})G(t-s/B_{N_{N}})ds\right].\label{three-three}
\end{eqnarray}

\section{\textbf{Theoretical examples}}

In this section, we show some examples of the likelihood function
constructed in the previous section, to estimate $\bold A,\textrm{ and }\bold B.$
Let $h(s)$ follow a quadratic exponential and $\bold B$ follow a)
a gamma function, and b) a logistic function. Infections in most of
the countries started declining after the availability of antiretroviral
therapies \cite{Conti2000,Hung2003}, and incidence in the recent
period was found to be stable in some countries like India \cite{Rao2009}.
This motivated us to choose a quadratic exponential to represent $h(s)$,
namely $h(s)=exp(\alpha_{1}s^{2}+\alpha_{2}s+\alpha_{3})$ for all
$-\infty<\alpha_{1},\alpha_{2},\alpha_{3}<\infty$. A quadratic exponential
function has been shown to be a good model for representing the above
declines in the incidence rates \cite{RaoASRS2005-1}. The incubation
period for AIDS is large as well as variable, therefore, functions
like the gamma, Weibull and logistic can mimic several shapes to fit
the incubation period data depending on their parameter values. Such
well-known functions were used by many researchers for modeling the
incubation period of AIDS. We now demonstrate the application of such
functions for the theory explained in section 2.

\subsection{Example 1: Gamma function. \textmd{If $\omega>0$ is the parameter
and $\Gamma(\omega)$ is the complete distribution function, then
the incomplete gamma distribution, $G(\omega;t_{j})=\frac{1}{\Gamma(\omega)}\int_{0}^{t_{j}}e^{-x}x^{\omega-1}dx,$
for $a\geq0,t_{j}\geq0$ and $a+t_{j}\neq0$}}

From the conditional probability equations from (\ref{2.5}) to (\ref{2.11}),
and the likelihood equations explained in the later part of section
3, the following are the likelihood equations without a drug and for
with three types of drugs: 

\begin{eqnarray}
L_{0}\left(\alpha_{1},\alpha_{2},\alpha_{3};\omega/P_{j}\right) & = & \prod_{j}a_{1}(j)a_{2}(j)-\prod_{j}a_{1}(j-1)a_{2}(j)\label{four-one}
\end{eqnarray}

where 
\begin{eqnarray*}
a_{1}(j) & = & \left[\int_{0}^{u_{j}}e^{\alpha_{1}s^{2}+\alpha_{2}s+\alpha_{3}}\left\{ \frac{1}{\Gamma\omega}\int_{0}^{t_{j}}e^{_{-(u_{j}-s)}}(u_{j}-s)^{\omega-1}du_{j}\right\} ds\right]^{T_{j}}\\
\\
a_{1}(j-1) & = & \left[\int_{0}^{u_{j-1}}e^{\alpha_{1}s^{2}+\alpha_{2}s+\alpha_{3}}\right.\\
 &  & \qquad\qquad\left.\times\left\{ \frac{1}{\Gamma\omega}\int_{0}^{t_{j-1}}e^{_{-(u_{j-1}-s)}}(u_{j-1}-s)^{\omega-1}du_{j-1}\right\} ds\right]^{T_{j}}\\
\\
a_{2}(j) & = & \left[\int_{0}^{u_{n}}e^{\alpha_{1}s^{2}+\alpha_{2}s+\alpha_{3}}\left\{ \frac{1}{\Gamma\omega}\int_{0}^{t_{n}}e^{_{-(u_{n}-s)}}(u_{n}-s)^{\omega-1}du_{n}\right\} ds\right]^{-T_{j}}
\end{eqnarray*}

\begin{eqnarray}
L_{1(2)}\left(\alpha_{1},\alpha_{2},\alpha_{3};\omega/P_{k}\right) & = & \prod_{k}a_{1}(k)a_{2}(k)-\prod_{k}a_{1}(k-1)a_{2}(k)\label{four-two}
\end{eqnarray}

where
\begin{eqnarray*}
a_{1}(k) & = & \left[\int_{0}^{u_{k}}e^{\alpha_{1}s^{2}+\alpha_{2}s+\alpha_{3}}\left\{ \frac{1}{\Gamma\omega}\int_{0}^{t_{k}}e^{_{-(u_{k}-s)}}(u_{k}-s)^{\omega-1}du_{k}\right\} ds\right]^{T_{k}}\\
\\
a_{1}(k-1) & = & \left[\int_{0}^{u_{k-1}}e^{\alpha_{1}s^{2}+\alpha_{2}s+\alpha_{3}}\right.{}\\
 &  & \qquad\qquad\left.\times\left\{ \frac{1}{\Gamma\omega}\int_{0}^{t_{k-1}}e^{_{-(u_{k-1}-s)}}(u_{k-1}-s)^{\omega-1}du_{k-1}\right\} ds\right]^{T_{k}}\\
\\
a_{2}(k) & = & \left[\int_{0}^{u_{n}}e^{\alpha_{1}s^{2}+\alpha_{2}s+\alpha_{3}}\left\{ \frac{1}{\Gamma\omega}\int_{0}^{t_{n}}e^{_{-(u_{n}-s)}}(u_{n}-s)^{\omega-1}du_{n}\right\} ds\right]^{-T_{k}}
\end{eqnarray*}

\begin{eqnarray}
L_{2(1)}\left(\alpha_{1},\alpha_{2},\alpha_{3};\omega/P_{l}\right) & = & \prod_{l}a_{1}(l)a_{2}(l)-\prod_{l}a_{1}(l-1)a_{2}(l)\label{four-three}
\end{eqnarray}

where

\begin{eqnarray*}
a_{1}(l) & = & \left[\int_{0}^{u_{l}}e^{\alpha_{1}s^{2}+\alpha_{2}s+\alpha_{3}}\left\{ \frac{1}{\Gamma\omega}\int_{0}^{t_{l}}e^{_{-(u_{l}-s)}}(u_{l}-s)^{\omega-1}du_{l}\right\} ds\right]^{T_{l}}\\
\\
a_{1}(l-1) & = & \left[\int_{0}^{u_{l-1}}e^{\alpha_{1}s^{2}+\alpha_{2}s+\alpha_{3}}\right.{}\\
 &  & \qquad\qquad\left.\times\left\{ \frac{1}{\Gamma\omega}\int_{0}^{t_{l-1}}e^{_{-(u_{l-1}-s)}}(u_{l-1}-s)^{\omega-1}du_{l-1}\right\} ds\right]^{T_{l}}\\
\\
a_{2}(l) & = & \left[\int_{0}^{u_{n}}e^{\alpha_{1}s^{2}+\alpha_{2}s+\alpha_{3}}\left\{ \frac{1}{\Gamma\omega}\int_{0}^{t_{n}}e^{_{-(u_{n}-s)}}(u_{n}-s)^{\omega-1}du_{n}\right\} ds\right]^{-T_{l}}
\end{eqnarray*}

\begin{eqnarray}
L_{1=2}\left(\alpha_{1},\alpha_{2},\alpha_{3};\omega/P_{p}\right) & = & \prod_{p}a_{1}(p)a_{2}(p)-\prod_{p}a_{1}(p-1)a_{2}(p)\label{four-f}
\end{eqnarray}

where

\begin{eqnarray*}
a_{1}(p) & = & \left[\int_{0}^{u_{p}}e^{\alpha_{1}s^{2}+\alpha_{2}s+\alpha_{3}}\left\{ \frac{1}{\Gamma\omega}\int_{0}^{t_{p}}e^{_{-(u_{p}-s)}}(u_{p}-s)^{\omega-1}du_{p}\right\} ds\right]^{T_{p}}\\
\\
a_{1}(p-1) & = & \left[\int_{0}^{u_{p-1}}e^{\alpha_{1}s^{2}+\alpha_{2}s+\alpha_{3}}\right.\\
 &  & \qquad\qquad\left.\times\left\{ \frac{1}{\Gamma\omega}\int_{0}^{t_{p-1}}e^{_{-(u_{p-1}-s)}}(u_{p-1}-s)^{\omega-1}du_{p-1}\right\} ds\right]^{T_{p}}\\
\\
a_{2}(p) & = & \left[\int_{0}^{u_{n}}e^{\alpha_{1}s^{2}+\alpha_{2}s+\alpha_{3}}\left\{ \frac{1}{\Gamma\omega}\int_{0}^{t_{n}}e^{_{-(u_{n}-s)}}(u_{n}-s)^{\omega-1}du_{n}\right\} ds\right]^{-T_{p}}
\end{eqnarray*}

\begin{eqnarray}
L_{3}\left(\alpha_{1},\alpha_{2},\alpha_{3};\omega/P_{m}\right) & = & \prod_{m}a_{1}(m)a_{2}(m)-\prod_{m}a_{1}(m-1)a_{2}(m)\label{four-five}
\end{eqnarray}

where

\begin{eqnarray*}
a_{1}(m) & = & \left[\int_{0}^{u_{m}}e^{\alpha_{1}s^{2}+\alpha_{2}s+\alpha_{3}}\left\{ \frac{1}{\Gamma\omega}\int_{0}^{t_{m}}e^{_{-(u_{m}-s)}}(u_{m}-s)^{\omega-1}du_{m}\right\} ds\right]^{T_{m}}\\
\\
a_{1}(m-1) & = & \left[\int_{0}^{u_{m-1}}e^{\alpha_{1}s^{2}+\alpha_{2}s+\alpha_{3}}\right.\\
 &  & \qquad\qquad\left.\times\left\{ \frac{1}{\Gamma\omega}\int_{0}^{t_{m-1}}e^{_{-(u_{m-1}-s)}}(u_{m-1}-s)^{\omega-1}du_{m-1}\right\} ds\right]^{T_{m}}\\
\\
a_{2}(m) & = & \left[\int_{0}^{u_{n}}e^{\alpha_{1}s^{2}+\alpha_{2}s+\alpha_{3}}\left\{ \frac{1}{\Gamma\omega}\int_{0}^{t_{n}}e^{_{-(u_{n}-s)}}(u_{n}-s)^{\omega-1}du_{n}\right\} ds\right]^{-T_{m}}
\end{eqnarray*}

\subsection{Example 2: Logistic function. \textmd{Suppose $\theta_{1},\theta_{2}$
are parameters and $F(\theta_{1},\theta_{2};t_{j})=\left\{ 1+e^{-(\frac{t_{j}-\theta_{1}}{\theta_{2}})}\right\} ^{-1},$
for $\theta_{1},\theta_{2}>0,$ is the distribution function. }}

The likelihood equations to obtain the parameters of logistic distribution
without drugs and for three types of drugs are as follows:

\begin{eqnarray}
L_{0}\left(\alpha_{1},\alpha_{2},\alpha_{3};\theta_{1},\theta_{2}/P_{j}\right) & = & \prod_{j}a'_{1}(j)a'_{2}(j)-\prod_{j}a'_{1}(j-1)a'_{2}(j)\label{4.2.1}
\end{eqnarray}

where

\begin{eqnarray*}
a'_{1}(j) & = & \left[\int_{0}^{u_{j}}e^{\alpha_{1}s^{2}+\alpha_{2}s+\alpha_{3}}\left\{ \left\{ 1+e^{-(\frac{u_{j}-\theta_{1}}{\theta_{2}})}\right\} ^{-1}\right\} ds\right]^{T_{j}}\\
\\
a'_{1}(j-1) & = & \left[\int_{0}^{u_{j-1}}e^{\alpha_{1}s^{2}+\alpha_{2}s+\alpha_{3}}\left\{ \left\{ 1+e^{-(\frac{u_{j-1}-\theta_{1}}{\theta_{2}})}\right\} ^{-1}\right\} ds\right]^{T_{j}}\\
\\
a'_{2}(j) & = & \left[\int_{0}^{u_{n}}e^{\alpha_{1}s^{2}+\alpha_{2}s+\alpha_{3}}\left\{ \left\{ 1+e^{-(\frac{u_{n}-\theta_{1}}{\theta_{2}})}\right\} ^{-1}\right\} ds\right]^{-T_{j}}
\end{eqnarray*}

\begin{eqnarray}
L_{1(2)}\left(\alpha_{1},\alpha_{2},\alpha_{3};\theta_{1},\theta_{2}/P_{k}\right) & = & \prod_{k}a'_{1}(k)a'_{2}(k)-\prod_{k}a'_{1}(k-1)a'_{2}(k)\label{4.2.2}
\end{eqnarray}

where

\begin{eqnarray*}
a'_{1}(k) & = & \left[\int_{0}^{u_{k}}e^{\alpha_{1}s^{2}+\alpha_{2}s+\alpha_{3}}\left\{ \left\{ 1+e^{-(\frac{u_{k}-\theta_{1}}{\theta_{2}})}\right\} ^{-1}\right\} ds\right]^{T_{k}}\\
\\
a'_{1}(k-1) & = & \left[\int_{0}^{u_{k-1}}e^{\alpha_{1}s^{2}+\alpha_{2}s+\alpha_{3}}\left\{ \left\{ 1+e^{-(\frac{u_{k-1}-\theta_{1}}{\theta_{2}})}\right\} ^{-1}\right\} ds\right]^{T_{k}}\\
\\
a'_{2}(k) & = & \left[\int_{0}^{u_{n}}e^{\alpha_{1}s^{2}+\alpha_{2}s+\alpha_{3}}\left\{ \left\{ 1+e^{-(\frac{u_{n}-\theta_{1}}{\theta_{2}})}\right\} ^{-1}\right\} ds\right]^{-T_{k}}
\end{eqnarray*}

\begin{eqnarray}
L_{2(1)}\left(\alpha_{1},\alpha_{2},\alpha_{3};\theta_{1},\theta_{2}/P_{l}\right) & = & \prod_{l}a'_{1}(l)a'_{2}(l)-\prod_{l}a'_{1}(l-1)a'_{2}(l)\label{4.2.3}
\end{eqnarray}

where

\begin{eqnarray*}
a'_{1}(l) & = & \left[\int_{0}^{u_{l}}e^{\alpha_{1}s^{2}+\alpha_{2}s+\alpha_{3}}\left\{ \left\{ 1+e^{-(\frac{u_{l}-\theta_{1}}{\theta_{2}})}\right\} ^{-1}\right\} ds\right]^{T_{l}}\\
\\
a'_{1}(l-1) & = & \left[\int_{0}^{u_{l-1}}e^{\alpha_{1}s^{2}+\alpha_{2}s+\alpha_{3}}\left\{ \left\{ 1+e^{-(\frac{u_{l-1}-\theta_{1}}{\theta_{2}})}\right\} ^{-1}\right\} ds\right]^{T_{l}}\\
\\
a'_{2}(l) & = & \left[\int_{0}^{u_{n}}e^{\alpha_{1}s^{2}+\alpha_{2}s+\alpha_{3}}\left\{ \left\{ 1+e^{-(\frac{u_{n}-\theta_{1}}{\theta_{2}})}\right\} ^{-1}\right\} ds\right]^{-T_{l}}
\end{eqnarray*}

\begin{eqnarray}
L_{1=2}\left(\alpha_{1},\alpha_{2},\alpha_{3};\theta_{1},\theta_{2}/P_{p}\right) & = & \prod_{p}a'_{1}(p)a'_{2}(p)-\prod_{p}a'_{1}(p-1)a'_{2}(p)\label{4.2.4}
\end{eqnarray}

where

\begin{eqnarray*}
a'_{1}(p) & = & \left[\int_{0}^{u_{p}}e^{\alpha_{1}s^{2}+\alpha_{2}s+\alpha_{3}}\left\{ \left\{ 1+e^{-(\frac{u_{p}-\theta_{1}}{\theta_{2}})}\right\} ^{-1}\right\} ds\right]^{T_{p}}\\
\\
a'_{1}(p-1) & = & \left[\int_{0}^{u_{p-1}}e^{\alpha_{1}s^{2}+\alpha_{2}s+\alpha_{3}}\left\{ \left\{ 1+e^{-(\frac{u_{p-1}-\theta_{1}}{\theta_{2}})}\right\} ^{-1}\right\} ds\right]^{T_{p}}\\
\\
a'_{2}(p) & = & \left[\int_{0}^{u_{n}}e^{\alpha_{1}s^{2}+\alpha_{2}s+\alpha_{3}}\left\{ \left\{ 1+e^{-(\frac{u_{n}-\theta_{1}}{\theta_{2}})}\right\} ^{-1}\right\} ds\right]^{-T_{p}}
\end{eqnarray*}

\begin{eqnarray}
L_{3}\left(\alpha_{1},\alpha_{2},\alpha_{3};\theta_{1},\theta_{2}/P_{m}\right) & = & \prod_{m}a'_{1}(m)a'_{2}(m)-\prod_{m}a'_{1}(m-1)a_{2}(m)\label{4.2.5}
\end{eqnarray}

where

\begin{eqnarray*}
a'_{1}(m) & = & \left[\int_{0}^{u_{m}}e^{\alpha_{1}s^{2}+\alpha_{2}s+\alpha_{3}}\left\{ \left\{ 1+e^{-(\frac{u_{m}-\theta_{1}}{\theta_{2}})}\right\} ^{-1}\right\} ds\right]^{T_{m}}\\
\\
a'_{1}(m-1) & = & \left[\int_{0}^{u_{m-1}}e^{\alpha_{1}s^{2}+\alpha_{2}s+\alpha_{3}}\left\{ \left\{ 1+e^{-(\frac{u_{m-1}-\theta_{1}}{\theta_{2}})}\right\} ^{-1}\right\} ds\right]^{T_{m}}\\
\\
a'_{2}(m) & = & \left[\int_{0}^{u_{n}}e^{\alpha_{1}s^{2}+\alpha_{2}s+\alpha_{3}}\left\{ \left\{ 1+e^{-(\frac{u_{n}-\theta_{1}}{\theta_{2}})}\right\} ^{-1}\right\} ds\right]^{-T_{m}}
\end{eqnarray*}

\subsection{Example 3: Log-normal function. \textmd{Suppose $\mu\textrm{ and }\sigma$
are parameters and $LNF(\mu,\sigma;t_{j})=\frac{1}{2}\left\{ 1+erf\left(\frac{\ln x-\mu}{\sigma\sqrt{2}}\right)\right\} ,$
for $\mu,\sigma>0,$ is the distribution function. (Here $erf\{.\}$
is the error function of the Gaussian function) }}

The likelihood equations to obtain the parameters of the logistic
distribution without drugs and for three types of drugs are as follows:

\begin{eqnarray}
L_{0}\left(\alpha_{1},\alpha_{2},\alpha_{3};\mu,\sigma/P_{j}\right) & = & \prod_{j}a''_{1}(j)a''_{2}(j)-\prod_{j}a''_{1}(j-1)a''_{2}(j)\label{4.2.1}
\end{eqnarray}

where

\begin{eqnarray*}
a''_{1}(j) & = & \left[\frac{1}{2}\int_{0}^{u_{j}}e^{\alpha_{1}s^{2}+\alpha_{2}s+\alpha_{3}}\left\{ 1+erf\left(\frac{\ln x-\mu}{\sigma\sqrt{2}}\right)\right\} ds\right]^{T_{j}}\\
\\
a''_{1}(j-1) & = & \left[\frac{1}{2}\int_{0}^{u_{j-1}}e^{\alpha_{1}s^{2}+\alpha_{2}s+\alpha_{3}}\left\{ 1+erf\left(\frac{\ln x-\mu}{\sigma\sqrt{2}}\right)\right\} ds\right]^{T_{j}}\\
\\
a''_{2}(j) & = & \left[\frac{1}{2}\int_{0}^{u_{n}}e^{\alpha_{1}s^{2}+\alpha_{2}s+\alpha_{3}}\left\{ 1+erf\left(\frac{\ln x-\mu}{\sigma\sqrt{2}}\right)\right\} ds\right]^{-T_{j}}
\end{eqnarray*}

\begin{eqnarray}
L_{1(2)}\left(\alpha_{1},\alpha_{2},\alpha_{3};\mu,\sigma/P_{k}\right) & = & \prod_{k}a''_{1}(k)a''_{2}(k)-\prod_{k}a''_{1}(k-1)a''_{2}(k)\label{4.2.2}
\end{eqnarray}

where

\begin{eqnarray*}
a''_{1}(k) & = & \left[\frac{1}{2}\int_{0}^{u_{k}}e^{\alpha_{1}s^{2}+\alpha_{2}s+\alpha_{3}}\left\{ 1+erf\left(\frac{\ln x-\mu}{\sigma\sqrt{2}}\right)\right\} ds\right]^{T_{k}}\\
\\
a''_{1}(k-1) & = & \left[\frac{1}{2}\int_{0}^{u_{k-1}}e^{\alpha_{1}s^{2}+\alpha_{2}s+\alpha_{3}}\left\{ 1+erf\left(\frac{\ln x-\mu}{\sigma\sqrt{2}}\right)\right\} ds\right]^{T_{k}}\\
\\
a''_{2}(k) & = & \left[\frac{1}{2}\int_{0}^{u_{n}}e^{\alpha_{1}s^{2}+\alpha_{2}s+\alpha_{3}}\left\{ 1+erf\left(\frac{\ln x-\mu}{\sigma\sqrt{2}}\right)\right\} ds\right]^{-T_{k}}
\end{eqnarray*}

\begin{eqnarray}
L_{2(1)}\left(\alpha_{1},\alpha_{2},\alpha_{3};\mu,\sigma/P_{l}\right) & = & \prod_{l}a''_{1}(l)a''_{2}(l)-\prod_{l}a''_{1}(l-1)a''_{2}(l)\label{4.2.3}
\end{eqnarray}

where

\begin{eqnarray*}
a''_{1}(l) & = & \left[\frac{1}{2}\int_{0}^{u_{l}}e^{\alpha_{1}s^{2}+\alpha_{2}s+\alpha_{3}}\left\{ 1+erf\left(\frac{\ln x-\mu}{\sigma\sqrt{2}}\right)\right\} ds\right]^{T_{l}}\\
\\
a''_{1}(l-1) & = & \left[\frac{1}{2}\int_{0}^{u_{l-1}}e^{\alpha_{1}s^{2}+\alpha_{2}s+\alpha_{3}}\left\{ 1+erf\left(\frac{\ln x-\mu}{\sigma\sqrt{2}}\right)\right\} ds\right]^{T_{l}}\\
\\
a''_{2}(l) & = & \left[\frac{1}{2}\int_{0}^{u_{n}}e^{\alpha_{1}s^{2}+\alpha_{2}s+\alpha_{3}}\left\{ 1+erf\left(\frac{\ln x-\mu}{\sigma\sqrt{2}}\right)\right\} ds\right]^{-T_{l}}
\end{eqnarray*}

\begin{eqnarray}
L_{1=2}\left(\alpha_{1},\alpha_{2},\alpha_{3};\mu,\sigma/P_{p}\right) & = & \prod_{p}a''_{1}(p)a''_{2}(p)-\prod_{p}a''_{1}(p-1)a''_{2}(p)\label{4.2.4}
\end{eqnarray}

where

\begin{eqnarray*}
a''_{1}(p) & = & \left[\frac{1}{2}\int_{0}^{u_{p}}e^{\alpha_{1}s^{2}+\alpha_{2}s+\alpha_{3}}\left\{ 1+erf\left(\frac{\ln x-\mu}{\sigma\sqrt{2}}\right)\right\} ds\right]^{T_{p}}\\
\\
a''_{1}(p-1) & = & \left[\frac{1}{2}\int_{0}^{u_{p-1}}e^{\alpha_{1}s^{2}+\alpha_{2}s+\alpha_{3}}\left\{ 1+erf\left(\frac{\ln x-\mu}{\sigma\sqrt{2}}\right)\right\} ds\right]^{T_{p}}\\
\\
a''_{2}(p) & = & \left[\frac{1}{2}\int_{0}^{u_{n}}e^{\alpha_{1}s^{2}+\alpha_{2}s+\alpha_{3}}\left\{ 1+erf\left(\frac{\ln x-\mu}{\sigma\sqrt{2}}\right)\right\} ds\right]^{-T_{p}}
\end{eqnarray*}

\begin{eqnarray}
L_{3}\left(\alpha_{1},\alpha_{2},\alpha_{3};\mu,\sigma/P_{m}\right) & = & \prod_{m}a'_{1}(m)a'_{2}(m)-\prod_{m}a'_{1}(m-1)a_{2}(m)\label{4.2.5}
\end{eqnarray}

where

\begin{eqnarray*}
a''_{1}(m) & = & \left[\frac{1}{2}\int_{0}^{u_{m}}e^{\alpha_{1}s^{2}+\alpha_{2}s+\alpha_{3}}\left\{ 1+erf\left(\frac{\ln x-\mu}{\sigma\sqrt{2}}\right)\right\} ds\right]^{T_{m}}\\
\\
a''_{1}(m-1) & = & \left[\frac{1}{2}\int_{0}^{u_{m-1}}e^{\alpha_{1}s^{2}+\alpha_{2}s+\alpha_{3}}\left\{ 1+erf\left(\frac{\ln x-\mu}{\sigma\sqrt{2}}\right)\right\} ds\right]^{T_{m}}\\
\\
a''_{2}(m) & = & \left[\frac{1}{2}\int_{0}^{u_{n}}e^{\alpha_{1}s^{2}+\alpha_{2}s+\alpha_{3}}\left\{ 1+erf\left(\frac{\ln x-\mu}{\sigma\sqrt{2}}\right)\right\} ds\right]^{-T_{m}}
\end{eqnarray*}

\section{\textbf{Age-structured populations}}

In this section we extend the models \ref{1.3} and \ref{1.4} to
accommodate age structure into the population mixing and epidemiology
parameters. The incubation period for children is shorter than that
of adults. Within the adult population there could be variability
due to age at the time of infection. There are studies that analyze
the HIV data on age collected at the time of infection to study parameters
like incubation period \cite{BECK2003}, and some studies incorporate
age structure in the models to explain the impact of an age-dependent
incubation period \cite{Griff2000}. Information on population age
structure is important source of data in a country with severe AIDS
epidemic. Countries with high number of young adults and with high-risk
behavior need special interventions in terms of behavioral counseling,
treatment of drugs, monitoring and evaluation of the epidemic. For
most of the countries with high numbers of HIV infected individuals,
age-related data for measuring impact of drugs are not available.
Virus transmission rates, disease progression rates and mortality
rates could be highly age-dependent. Improving surveillance activities
by age-structure of the HIV infected and susceptible populations would
benefit the overall disease control programs in a country. In the
absence of availability of cohort data, the methods explained in section
2 could be of great use to estimate the incubation period. The analysis
and method explained there could be carried out based on the data
available for individuals of every age (rounded to closest integer).
We describe the age-structure model and the method to obtain the incubation
period in this section by considering $j$ age groups. In a hospital
set-up it is relatively easy to follow cohorts of age groups compared
to following cohorts of individuals for each age group. 

Suppose the population in the $j^{th}$ age group is divided into
$X_{j}$ susceptible, $Y_{0,j}$, $Y_{1,j}$, $Y_{2,j}$, $Y_{3,j}$
are infected and $D_{z_{0},j}$, $D_{z_{1},j}$, $D_{z_{2},j}$, $D_{z_{3},j}$
individuals with the disease without drugs, and for \emph{drug1, drug2,
drug3} respectively. The differential equations explaining these variables
are 

\begin{eqnarray}
\frac{dX_{j}}{dt} & = & \Lambda_{j}-\left(\lambda_{jk}^{0}+\lambda_{jk}^{1}+\lambda_{jk}^{2}+\lambda_{jk}^{3}+\mu_{j}\right)X_{j},\nonumber \\
\frac{dY_{0,j}}{dt} & = & \lambda_{jk}^{0}X_{j}-\left\{ \left(\int_{\mathbb{R}}z_{0,j}dG(z_{0,j})\right)^{-1}+\mu_{j}\right\} Y_{0,j},\nonumber \\
\frac{dY_{1,j}}{dt} & = & \lambda_{jk}^{1}X_{j}-\left\{ \left(\int_{\mathbb{R}}z_{1,j}dG(z_{1,j})\right)^{-1}+\mu_{j}\right\} Y_{1,j},\nonumber \\
\frac{dY_{2,j}}{dt} & = & \lambda_{jk}^{2}X_{j}-\left\{ \left(\int_{\mathbb{R}}z_{2,j}dG(z_{2,j})\right)^{-1}+\mu_{j}\right\} Y_{2,j},\nonumber \\
\frac{dY_{3,j}}{dt} & = & \lambda_{jk}^{3}X_{j}-\left\{ \left(\int_{\mathbb{R}}z_{3,j}dG(z_{3,j})\right)^{-1}+\mu_{j}\right\} Y_{3,j},\nonumber \\
\frac{dD_{z_{0},j}}{dt} & = & \left(\int_{\mathbb{R}}z_{0,j}dG(z_{0,j})\right)^{-1}Y_{0,j}-\gamma_{0,j}D_{z_{0},j}\nonumber \\
\frac{dD_{z_{1},j}}{dt} & = & \left(\int_{\mathbb{R}}z_{1,j}dG(z_{1,j})\right)^{-1}Y_{1,j}-\gamma_{1,j}D_{z_{1},j}\label{6.1}\\
\frac{dD_{z_{2},j}}{dt} & = & \left(\int_{\mathbb{R}}z_{2,j}dF(z_{2,j})\right)^{-1}Y_{2,j}-\gamma_{2,j}D_{z_{2},j}\nonumber \\
\frac{dD_{z_{3},j}}{dt} & = & \left(\int_{\mathbb{R}}z_{3,j}dG(z_{3,j})\right)^{-1}Y_{3,j}-\gamma_{3,j}D_{z_{3},j}.\nonumber 
\end{eqnarray}

Here, $\Lambda_{j}$ is the input of susceptibles for the individuals
in the age group $j$, $\mu_{j}$ is the mortality rate, $\lambda_{jk}^{0}$,$\lambda_{jk}^{1}$
,$\lambda_{jk}^{2}$ and $\lambda_{jk}^{3}$ are the forces of infection
at which a susceptible in the age group $j$ is infected by an infected
individual in the age group $k$ and$\gamma_{0,j}$, $\gamma_{1,j}$,
$\gamma_{2,j}$ and $\gamma_{3,j}$ are disease related mortality
rates for the infected individuals in the age group $j$ without drugs,
and with \emph{drug1, drug2, and drug3} for the individuals. $\left(\int_{\mathbb{R}}z_{i,j}dG(z_{i,j})\right)^{-1}$
is the rate of disease progression for the infected individual for
the age group $j$ for the drug type $i.$ Special attention is necessary
in data collection for understanding the forces of infection by age
group.

If there are $n$ drug types available then the general model describing
the dynamics of various variables described above is as follows:

\begin{eqnarray}
\frac{dX_{j}}{dt} & = & \Lambda_{j}-\left(\lambda_{jk}^{0}+\lambda_{jk}^{1}+\lambda_{jk}^{2}+\lambda_{jk}^{3}+\mu_{j}\right)X_{j},\nonumber \\
\frac{dY_{0,j}}{dt} & = & \lambda_{jk}^{0}X_{j}-\left\{ \left(\int_{\mathbb{R}}z_{0,j}dG(z_{0,j})\right)^{-1}+\mu_{j}\right\} Y_{0,j},\nonumber \\
\vdots &  & \vdots\nonumber \\
\vdots &  & \vdots\nonumber \\
\frac{dY_{n,j}}{dt} & = & \lambda_{jk}^{n}X_{j}-\left\{ \left(\int_{\mathbb{R}}z_{n,j}dG(z_{n,j})\right)^{-1}+\mu_{j}\right\} Y_{n,j},\nonumber \\
\frac{dD_{z_{0},j}}{dt} & = & \left(\int_{\mathbb{R}}z_{0,j}dG(z_{0,j})\right)^{-1}Y_{0,j}-\gamma_{0,j}D_{z_{0},j},\nonumber \\
\vdots &  & \vdots\label{6.2}\\
\vdots &  & \vdots\nonumber \\
\frac{dD_{z_{n},j}}{dt} & = & \left(\int_{\mathbb{R}}z_{n,j}dG(z_{n,j})\right)^{-1}Y_{n,j}-\gamma_{n,j}D_{z_{n},j}.\nonumber 
\end{eqnarray}

where $\alpha_{i,j}$ is the mortality rate of infected individuals
of drug type $i$ in the age group $j.$

\subsection{Varying incubation periods for age-structured populations}

We are interested in the average incubation period for a group of
individuals in the age group $j$. If $H(j)$ is the time of infection
and $Z(j)$ is the incubation period for the $j^{th}$ age group,
then the time of onset of the disease for this age group is $D(j)=H(j)+Z(j).$
This is the time of onset of the disease for an individual who acquired
the infection while in the $j^{th}$ age group. Development of the
disease will be some time units (for example: months, years) after
infection at age $j.$ An individual who acquired the infection at
age $j$ is assumed to develop the full disease before completion
of the same age $j$ or $>j$. Given $H(j)$, for some $j$, then
$D(j)$ is allowed to occur at age $j'$($j'=j,j+1,\cdots,j+\omega,$
where $j+\omega$ is the last age group for the possibility of infection).
Clearly, $H(j)\leq D(j).$ $H(j)=D(j)$ is possible if an individual
acquired infection and attains disease before completion of age $j.$
One can do analysis using a bi-annual (or half-yearly) aging process.

\begin{figure}
\includegraphics[scale=0.6]{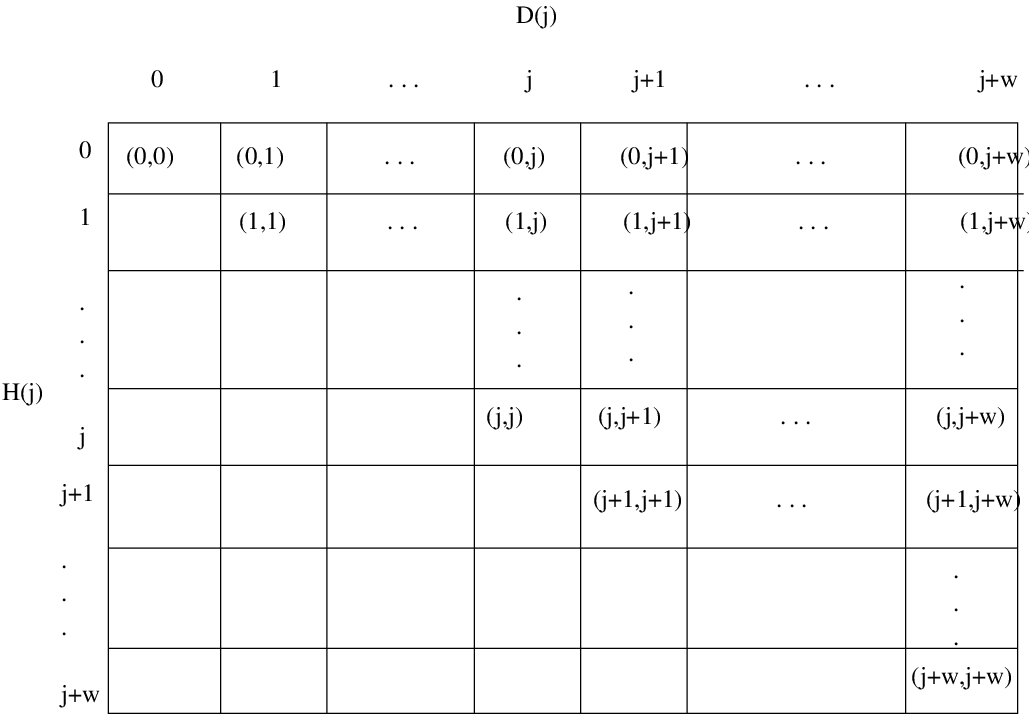}

\caption{Age-structured infection and disease development matrix. Here row
values indicate infection age group ($H(j)$) and column values for
age group in which infected individual developed disease ($D(j)$).
An individual who acquired the infection in $j$, and developed disease
in $j+\omega$, is indicated by the cell ($j,j+\omega$). }
\label{fig-ages}

\end{figure}

Consider an infection and disease development matrix (see figure \ref{fig-ages})
where each cell $(j,j')$ denotes the (infection age groups, disease
onset age groups) for $j=0,1,2,...,j+\omega$; $j'=0,1,2,...,j+\omega.$
Only those cells for which $j\leq j'$ are provided, and other cells
are left blank for which the incubation period is not defined. In
the matrix, all the eligible cells are denoted, so obviously there
are more cells present where the condition $j\leq j'$ is satisfied,
and also $j$ is very low. (In fact, the average incubation period
is not beyond a certain duration. It is not intended in the matrix
to suggest that the lower the value of $j$ then the larger the value
of incubation period). If the age of infection is higher, for some
$j$, and towards the last few possible age groups, then it is possible
that $j'-j$ is shorter because individuals die naturally in old age.
At the same time, the chance of infection in the very higher age groups
(say 60+) is negligible for HIV (unless there are some rare causes).
In the absence of age specific cohorts of infected individuals and
follow-up data, it is not feasible to calculate disease progression
rates and survival probabilities using direct cohort methods. In this
section, we extend the method given in section 2 to estimate the average
disease progression rates (or average incubation periods) for infections
in age group $j$. This method is dependent on infection densities
and data on disease occurrences for the age group $j.$

Let $p(t,j)$ and $q(t,j)$ be the probability density functions of
infection density and incubation period for the age group $j.$ If
$Q(t,j)$ is the distribution function of the incubation period, then
$Q(t,j)=\int_{-\infty}^{t}q(t,j)dt.$ Now, the convolution of $p(t,j)$
and $Q(t,j)$ is given by 

\begin{eqnarray*}
C(s,t) & = & \int_{-\infty}^{\infty}p(t,j)Q(t-s,j)ds.
\end{eqnarray*}

We call $C$ the convolution of $p$ and $Q$ (i.e. $p*Q$, where
$*$ is the convolution operator). Therefore,

\begin{eqnarray*}
p*Q & = & \int_{-\infty}^{\infty}p(t,j)Q(t-s,j)ds.
\end{eqnarray*}

Suppose an individual is diagnosed with a disease at age $j$ in the
year $U_{k}.$ Then there is a possibility that this individual acquired
the infection in any of the years prior to $U_{k}$ (provided this
individual is born in the year $\geq U_{0}$). Similarly, all those
individuals who are diagnosed with the disease at age $j+w$ in the
year $U_{n}$ have actually acquired infection in any of the years
from $U_{0}$ to $U_{n}.$ In the same way, an individual infected
at age $j$ will be diagnosed with the disease in an age group $\geq j.$
We consider model (\ref{6.1}), where four types of drugs were considered
in section 2.

Let $A_{0}(j),$$A_{1}(j),$ $A_{2}(j),$$A_{3}(j)$ be the parameter
sets in age group $j$ for the four kind of drugs. Let $B_{0}(j),$$B_{1}(j),$
$B_{2}(j),$$B_{3}(j)$ be the parameter sets $C$ and $E_{0}(j),$$E_{1}(j),$
$E_{2}(j),$$E_{3}(j)$ be the corresponding events of diagnosis of
disease in the age group $j$ for the four types of drugs. The cumulative
number of diagnosed disease cases up to $U_{n}$ for individuals who
are diagnosed in the age group $j$ is 

\begin{eqnarray*}
J(U_{0}<s<U_{n},j) & = & \sum_{j*=0}^{j}I(j*,j),
\end{eqnarray*}

where $I(0,j),I(1,j),...,I(j,j)$ are the numbers of disease cases
diagnosed in age group $j$, and acquired the infection in the age
group $0,1,...,j.$ 

\begin{eqnarray*}
I(0,j) & = & \int_{0}^{U_{n-k}}p(t,0)Q(t-s,j)ds+\int_{U_{n-k}}^{U_{n-l}}p(t,0)Q(t-s,j)ds\\
\\
 &  & +\int_{U_{n-l}}^{U_{n-m}}p(t,0)Q(t-s,j)ds+\int_{U_{n-m}}^{U_{n}}p(t,0)Q(t-s,j)ds
\end{eqnarray*}

\begin{eqnarray*}
 & = & \int_{0}^{U_{n-k}}p(t,0/A_{0})Q(t-s,j/B_{0})ds+\int_{U_{n-k}}^{U_{n-l}}p(t,0/A_{1})Q(t-s,j/B_{1})ds\\
\\
 &  & +\int_{U_{n-l}}^{U_{n-m}}p(t,0/A_{2})Q(t-s,j/B_{2})ds+\int_{U_{n-m}}^{U_{n}}p(t,0/A_{3})Q(t-s,j/B_{3})ds
\end{eqnarray*}

\begin{eqnarray*}
I(1,j) & = & \int_{0}^{U_{n-k}}p(t,1)Q(t-s,j)ds+\int_{U_{n-k}}^{U_{n-l}}p(t,1)Q(t-s,j)ds\\
\\
 &  & +\int_{U_{n-l}}^{U_{n-m}}p(t,1)Q(t-s,j)ds+\int_{U_{n-m}}^{U_{n}}p(t,1)Q(t-s,j)ds
\end{eqnarray*}

\begin{eqnarray*}
 & = & \int_{0}^{U_{n-k}}p(t,1/A_{0})Q(t-s,j/B_{0})ds+\int_{U_{n-k}}^{U_{n-l}}p(t,1/A_{1})Q(t-s,j/B_{1})ds\\
\\
 &  & +\int_{U_{n-l}}^{U_{n-m}}p(t,1/A_{2})Q(t-s,j/B_{2})ds+\int_{U_{n-m}}^{U_{n}}p(t,1/A_{3})Q(t-s,j/B_{3})ds
\end{eqnarray*}

\begin{eqnarray*}
 & \vdots
\end{eqnarray*}

\begin{eqnarray*}
I(j,j) & = & \int_{0}^{U_{n-k}}p(t,j)Q(t-s,j)ds+\int_{U_{n-k}}^{U_{n-l}}p(t,j)Q(t-s,j)ds\\
\\
 &  & +\int_{U_{n-l}}^{U_{n-m}}p(t,j)Q(t-s,j)ds+\int_{U_{n-m}}^{U_{n}}p(t,j)Q(t-s,j)ds
\end{eqnarray*}

\begin{eqnarray*}
 & = & \int_{0}^{U_{n-k}}p(t,j/A_{0})Q(t-s,j/B_{0})ds+\int_{U_{n-k}}^{U_{n-l}}p(t,j/A_{1})Q(t-s,j/B_{1})ds\\
\\
 &  & +\int_{U_{n-l}}^{U_{n-m}}p(t,j/A_{2})Q(t-s,j/B_{2})ds+\int_{U_{n-m}}^{U_{n}}p(t,j/A_{3})Q(t-s,j/B_{3})ds
\end{eqnarray*}

Similarly for unstructured populations, we assume that $U_{n-k}$
is the time of the introduction of drugs after the first year of detection
of the disease in $U_{0}.$ If $E_{1}(j)\in[U_{n-k},U_{n-l})$ and
$E_{2}(j)\in[U_{n-l},U_{n-m}),$ then $Z_{1}(j)<Z_{2}(j),$ otherwise
if $E_{2}(j)\in[U_{n-k},U_{n-l})$ and $E_{1}(j)\in[U_{n-l},U_{n-m})$
then $Z_{2}(j)<Z_{1}(j).$ If $E_{1}(j),E_{2}(j)\in[U_{n-k},U_{n-m})$,
then $Z_{1}(j)=Z_{2}(j).$ Given an individual who was diagnosed with
the disease in the age group $j$ before $U_{n}$ is already developed
in one of the four intervals $[U_{0},U_{n-k})$, $[U_{n-k},U_{n-l})$,
$[U_{n-l},U_{n-m})$ and $[U_{n-m},U_{n}).$ If $E_{0}(j)\in[U_{i'-1},U_{i'})\subseteq[U_{0},U_{n-k})$,
(for drug type $i'$), then the conditional probability of occurrence
of $E_{0}(j)$ given $E(j)$ is 

\begin{eqnarray*}
Pr\left[E_{0}(j)/E(j)\right] & = & Pr\left[U_{i'-1}\leq J\leq U_{i'},j/J\leq U_{n}\right]\\
\\
 & = & \frac{J\left[A_{0}(j),B_{0}(j)/U_{i'},j\right]-J\left[A_{0}(j),B_{0}(j)/u_{i'-1},j\right]}{J\left[A_{0}(j),B_{0}(j)/U_{n},j\right]},
\end{eqnarray*}

where $J$ values for $E_{0}(j)$ are given by:

\begin{eqnarray*}
J\left[A_{0}(j),B_{0}(j)/U_{i'},j\right] & = & \int_{0}^{U_{i'}}p(t,0/A_{0})Q(t-s,j/B_{0})ds\\
\\
+\int_{0}^{U_{i'}}p(t,1/A_{0})Q(t-s,j/B_{0})ds & \cdots & +\int_{0}^{U_{i'}}p(t,j/A_{0})Q(t-s,j/B_{0})ds
\end{eqnarray*}

\begin{eqnarray*}
J\left[A_{0}(j),B_{0}(j)/U_{i'-1},j\right] & = & \int_{0}^{U_{i'-1}}p(t,0/A_{0})Q(t-s,j/B_{0})ds\\
\\
+\int_{0}^{U_{i'-1}}p(t,1/A_{0})Q(t-s,j/B_{0})ds & \cdots & +\int_{0}^{U_{i'-1}}p(t,j/A_{0})Q(t-s,j/B_{0})ds
\end{eqnarray*}

\begin{eqnarray*}
J\left[A_{1}(j),B_{1}(j)/U_{n},j\right] & = & \int_{0}^{U_{n}}p(t,0/A_{0})Q(t-s,j/B_{0})ds\\
\\
+\int_{0}^{U_{n}}p(t,1/A_{0})Q(t-s,j/B_{0})ds & \cdots & +\int_{0}^{U_{n}}p(t,j/A_{0})Q(t-s,j/B_{0})ds
\end{eqnarray*}

The above probability expressions are for the case without drug interventions.
When \emph{drugs} were initiated at $U_{n-k},$ then these probabilities
changed according to the occurrence of $E_{1}(j),E_{2}(j),E_{3}(j).$
Suppose $E_{1}(j)\cap E_{2}(j)=\emptyset.$ If $[U_{k'-1},U_{k'})\subseteq[U_{n-k},U_{n-l})$,
and $E_{1}\in[U_{n-k},U_{n-l}),$ then 

\begin{eqnarray*}
Pr\left[E_{1}(j)/E(j)\right] & = & Pr\left[U_{k'-1}\leq J\leq U_{k'},j/J\leq U_{n}\right]\\
\\
 & = & \frac{J\left[A_{1}(j),B_{1}(j)/U_{k'},j\right]-J\left[A_{1}(j),B_{1}(j)/u_{k'-1},j\right]}{J\left[A_{1}(j),B_{1}(j)/U_{n},j\right]},
\end{eqnarray*}

where $J$ values for $E_{1}(j)$ are given by

\begin{eqnarray*}
J\left[A_{1}(j),B_{1}(j)/U_{k'},j\right] & = & \int_{0}^{U_{k'}}p(t,0/A_{1})Q(t-s,j/B_{1})ds\\
\\
+\int_{0}^{U_{k'}}p(t,1/A_{1})Q(t-s,j/B_{1})ds & \cdots & +\int_{0}^{U_{k'}}p(t,j/A_{1})Q(t-s,j/B_{1})ds
\end{eqnarray*}

\begin{eqnarray*}
J\left[A_{1}(j),B_{1}(j)/U_{k'-1},j\right] & = & \int_{0}^{U_{k'-1}}p(t,0/A_{1})Q(t-s,j/B_{1})ds\\
\\
+\int_{0}^{U_{k'-1}}p(t,1/A_{1})Q(t-s,j/B_{1})ds & \cdots & +\int_{0}^{U_{k'-1}}p(t,j/A_{1})Q(t-s,j/B_{1})ds
\end{eqnarray*}

\begin{eqnarray*}
J\left[A_{1}(j),B_{1}(j)/U_{n},j\right] & = & \int_{0}^{U_{n}}p(t,0/A_{1})Q(t-s,j/B_{1})ds\\
\\
+\int_{0}^{U_{n}}p(t,1/A_{1})Q(t-s,j/B_{1})ds & \cdots & +\int_{0}^{U_{n}}p(t,j/A_{1})Q(t-s,j/B_{1})ds
\end{eqnarray*}

In the above, instead of $E_{1}(j),$ if $E_{2}\in[U_{n-k},U_{n-l})$,
then the probabilities would be

\begin{eqnarray*}
Pr\left[E_{2}(j)/E(j)\right] & = & Pr\left[U_{k'-1}\leq J\leq U_{k'},j/J\leq U_{n}\right]\\
\\
 & = & \frac{J\left[A_{2}(j),B_{2}(j)/U_{k'},j\right]-J\left[A_{2}(j),B_{2}(j)/u_{k'-1},j\right]}{J\left[A_{2}(j),B_{2}(j)/U_{n},j\right]},
\end{eqnarray*}

where $J$ values for $E_{1}(j)$ are given by

\begin{eqnarray*}
J\left[A_{2}(j),B_{2}(j)/U_{k'},j\right] & = & \int_{0}^{U_{k'}}p(t,0/A_{2})Q(t-s,j/B_{2})ds\\
\\
+\int_{0}^{U_{k'}}p(t,1/A_{2})Q(t-s,j/B_{2})ds & \cdots & +\int_{0}^{U_{k'}}p(t,j/A_{2})Q(t-s,j/B_{2})ds
\end{eqnarray*}

\begin{eqnarray*}
J\left[A_{2}(j),B_{2}(j)/U_{k'-1},j\right] & = & \int_{0}^{U_{k'-1}}p(t,0/A_{2})Q(t-s,j/B_{2})ds\\
\\
+\int_{0}^{U_{k'-1}}p(t,1/A_{2})Q(t-s,j/B_{2})ds & \cdots & +\int_{0}^{U_{k'-1}}p(t,j/A_{2})Q(t-s,j/B_{2})ds
\end{eqnarray*}

\begin{eqnarray*}
J\left[A_{2}(j),B_{2}(j)/U_{n},j\right] & = & \int_{0}^{U_{n}}p(t,0/A_{2})Q(t-s,j/B_{2})ds\\
\\
+\int_{0}^{U_{n}}p(t,1/A_{2})Q(t-s,j/B_{2})ds &  & +\int_{0}^{U_{n}}p(t,j/A_{2})Q(t-s,j/B_{2})ds
\end{eqnarray*}

If $[U_{l-1},U_{l})\subseteq[U_{n-l},U_{n-m})$, and $E_{1}\in[U_{n-l},U_{n-m}),$
then

\begin{eqnarray*}
Pr\left[E_{1}(j)/E(j)\right] & = & Pr\left[U_{l'-1}\leq J\leq U_{l'},j/J\leq U_{n}\right]\\
\\
 & = & \frac{J\left[A_{1}(j),B_{1}(j)/U_{l'},j\right]-J\left[A_{1}(j),B_{1}(j)/u_{l'-1},j\right]}{J\left[A_{1}(j),B_{1}(j)/U_{n},j\right]},
\end{eqnarray*}

where $J$ values for $E_{1}(j)$ are given by

\begin{eqnarray*}
J\left[A_{1}(j),B_{1}(j)/U_{l'},j\right] & = & \int_{0}^{U_{l'}}p(t,0/A_{1})Q(t-s,j/B_{1})ds\\
\\
+\int_{0}^{U_{l'}}p(t,1/A_{1})Q(t-s,j/B_{1})ds & \cdots & +\int_{0}^{U_{l'}}p(t,j/A_{1})Q(t-s,j/B_{1})ds
\end{eqnarray*}

\begin{eqnarray*}
J\left[A_{1}(j),B_{1}(j)/U_{l'-1},j\right] & = & \int_{0}^{U_{l'-1}}p(t,0/A_{1})Q(t-s,j/B_{1})ds\\
\\
+\int_{0}^{U_{l'-1}}p(t,1/A_{1})Q(t-s,j/B_{1})ds & \cdots & +\int_{0}^{U_{l'-1}}p(t,j/A_{1})Q(t-s,j/B_{1})ds
\end{eqnarray*}

\begin{eqnarray*}
J\left[A_{1}(j),B_{1}(j)/U_{n},j\right] & = & \int_{0}^{U_{n}}p(t,0/A_{1})Q(t-s,j/B_{1})ds\\
\\
+\int_{0}^{U_{n}}p(t,1/A_{1})Q(t-s,j/B_{1})ds & \cdots & +\int_{0}^{U_{n}}p(t,j/A_{1})Q(t-s,j/B_{1})ds
\end{eqnarray*}

Suppose $[U_{l-1},U_{l})\subseteq[U_{n-l},U_{n-m})$, and $E_{2}\in[U_{n-l},U_{n-m})$
then

\begin{eqnarray*}
Pr\left[E_{2}(j)/E(j)\right] & = & Pr\left[U_{l'-1}\leq J\leq U_{l'},j/J\leq U_{n}\right]\\
\\
 & = & \frac{J\left[A_{2}(j),B_{2}(j)/U_{l'},j\right]-J\left[A_{2}(j),B_{2}(j)/u_{l'-1},j\right]}{J\left[A_{2}(j),B_{2}(j)/U_{n},j\right]}
\end{eqnarray*}

where $J$ values for $E_{2}(j)$ are given as below:

\begin{eqnarray*}
J\left[A_{2}(j),B_{2}(j)/U_{l'},j\right] & = & \int_{0}^{U_{l'}}p(t,0/A_{2})Q(t-s,j/B_{2})ds\\
\\
+\int_{0}^{U_{l'}}p(t,1/A_{2})Q(t-s,j/B_{2})ds & \cdots & +\int_{0}^{U_{l'}}p(t,j/A_{2})Q(t-s,j/B_{2})ds
\end{eqnarray*}

\begin{eqnarray*}
J\left[A_{2}(j),B_{2}(j)/U_{l'-1},j\right] & = & \int_{0}^{U_{l'-1}}p(t,0/A_{2})Q(t-s,j/B_{2})ds\\
\\
+\int_{0}^{U_{l'-1}}p(t,1/A_{2})Q(t-s,j/B_{2})ds & \cdots & +\int_{0}^{U_{l'-1}}p(t,j/A_{2})Q(t-s,j/B_{2})ds
\end{eqnarray*}

\begin{eqnarray*}
J\left[A_{2}(j),B_{2}(j)/U_{n},j\right] & = & \int_{0}^{U_{n}}p(t,0/A_{2})Q(t-s,j/B_{2})ds\\
\\
+\int_{0}^{U_{n}}p(t,1/A_{2})Q(t-s,j/B_{2})ds & \cdots & +\int_{0}^{U_{n}}p(t,j/A_{2})Q(t-s,j/B_{2})ds
\end{eqnarray*}

If $E_{1}(j)=E_{2}(j)\in[U_{p'-1},U_{p'})\subseteq[U_{n-k},U_{n-m})$
i.e. $Z_{1}(j)=Z_{2}(j),$ then the conditional probabilities contain
the same parameter sets. The probabilities for this situation are

\begin{eqnarray*}
Pr\left[E_{1}(j)=E_{2}(j)/E(j)\right] & = & Pr\left[U_{p'-1}\leq J\leq U_{p'},j/J\leq U_{n}\right]\\
\\
 & = & \frac{J\left[A_{2}(j),B_{2}(j)/U_{p'},j\right]-J\left[A_{2}(j),B_{2}(j)/u_{p'-1},j\right]}{J\left[A_{2}(j),B_{2}(j)/U_{n},j\right]},
\end{eqnarray*}

where $J$ values for $E_{2}(j)$ are given by

\begin{eqnarray*}
J\left[A_{2}(j),B_{2}(j)/U_{p'},j\right] & = & \int_{0}^{U_{p'}}p(t,0/A_{2})Q(t-s,j/B_{2})ds\\
\\
+\int_{0}^{U_{p'}}p(t,1/A_{2})Q(t-s,j/B_{2})ds & \cdots & +\int_{0}^{U_{p'}}p(t,j/A_{2})Q(t-s,j/B_{2})ds
\end{eqnarray*}

\begin{eqnarray*}
J\left[A_{2}(j),B_{2}(j)/U_{p'-1},j\right] & = & \int_{0}^{U_{p'-1}}p(t,0/A_{2})Q(t-s,j/B_{2})ds\\
\\
+\int_{0}^{U_{p'-1}}p(t,1/A_{2})Q(t-s,j/B_{2})ds & \cdots & +\int_{0}^{U_{p'-1}}p(t,j/A_{2})Q(t-s,j/B_{2})ds
\end{eqnarray*}

\begin{eqnarray*}
J\left[A_{2}(j),B_{2}(j)/U_{n},j\right] & = & \int_{0}^{U_{n}}p(t,0/A_{2})Q(t-s,j/B_{2})ds\\
\\
+\int_{0}^{U_{n}}p(t,1/A_{2})Q(t-s,j/B_{2})ds & \cdots & +\int_{0}^{U_{n}}p(t,j/A_{2})Q(t-s,j/B_{2})ds
\end{eqnarray*}

Since $Z_{3}(j)>\left\{ Z_{0}(j),Z_{1}(j),Z_{2}(j)\right\} $, suppose
$E_{3}(j)\in[U_{m'-1},U_{m'})\subseteq[U_{n-m},U_{n}]$, now above
probabilities are

\begin{eqnarray*}
Pr\left[E_{3}(j)/E(j)\right] & = & Pr\left[U_{m'-1}\leq J\leq U_{m'},j/J\leq U_{n}\right]\\
\\
 & = & \frac{J\left[A_{3}(j),B_{3}(j)/U_{m'},j\right]-J\left[A_{3}(j),B_{3}(j)/u_{m'-1},j\right]}{J\left[A_{3}(j),B_{3}(j)/U_{n},j\right]}
\end{eqnarray*}

where $J$ values for $E_{3}(j)$ are given by
\begin{eqnarray*}
J\left[A_{3}(j),B_{3}(j)/U_{m'},j\right] & = & \int_{0}^{U_{m'}}p(t,0/A_{3})Q(t-s,j/B_{3})ds\\
\\
+\int_{0}^{U_{m'}}p(t,1/A_{3})Q(t-s,j/B_{3})ds & \cdots & +\int_{0}^{U_{m'}}p(t,j/A_{3})Q(t-s,j/B_{3})ds
\end{eqnarray*}

\begin{eqnarray*}
J\left[A_{3}(j),B_{3}(j)/U_{m'-1},j\right] & = & \int_{0}^{U_{m'-1}}p(t,0/A_{3})Q(t-s,j/B_{3})ds\\
\\
+\int_{0}^{U_{m'-1}}p(t,1/A_{3})Q(t-s,j/B_{3})ds & \cdots & +\int_{0}^{U_{m'-1}}p(t,j/A_{3})Q(t-s,j/B_{3})ds
\end{eqnarray*}

\begin{eqnarray*}
J\left[A_{3}(j),B_{3}(j)/U_{n},j\right] & = & \int_{0}^{U_{n}}p(t,0/A_{3})Q(t-s,j/B_{3})ds\\
\\
+\int_{0}^{U_{n}}p(t,1/A_{3})Q(t-s,j/B_{3})ds & \cdots & +\int_{0}^{U_{n}}p(t,j/A_{3})Q(t-s,j/B_{3})ds
\end{eqnarray*}

Using the above conditional probabilities, likelihood functions are
constructed by assuming some parametric form for the diagnosed disease
cases. For each age group above, analysis is conducted to estimate
the incubation periods by age group.

\section{\textbf{Conclusions}}

The methods and models developed support further biological and epidemiological
experiments in the HIV infected population. As per the current WHO
guidelines, ART is prescribed only when CD4 count reaches 250. Experiments
indicate mortality rate among HIV infected population drops after
individuals are on ART, and hence expected life years remaining once
individuals reaches CD4 = 250 is different for those individuals who
are on ART and who are not on ART. After providing ART for all the
eligible people, the length of life gained by individuals can be measured
and resultant functional form can be modeled. Similarly, a model can
be built for lengths of lives for those individuals who reach CD4=250
and not on ART. 

\begin{figure}
\includegraphics[scale=0.6]{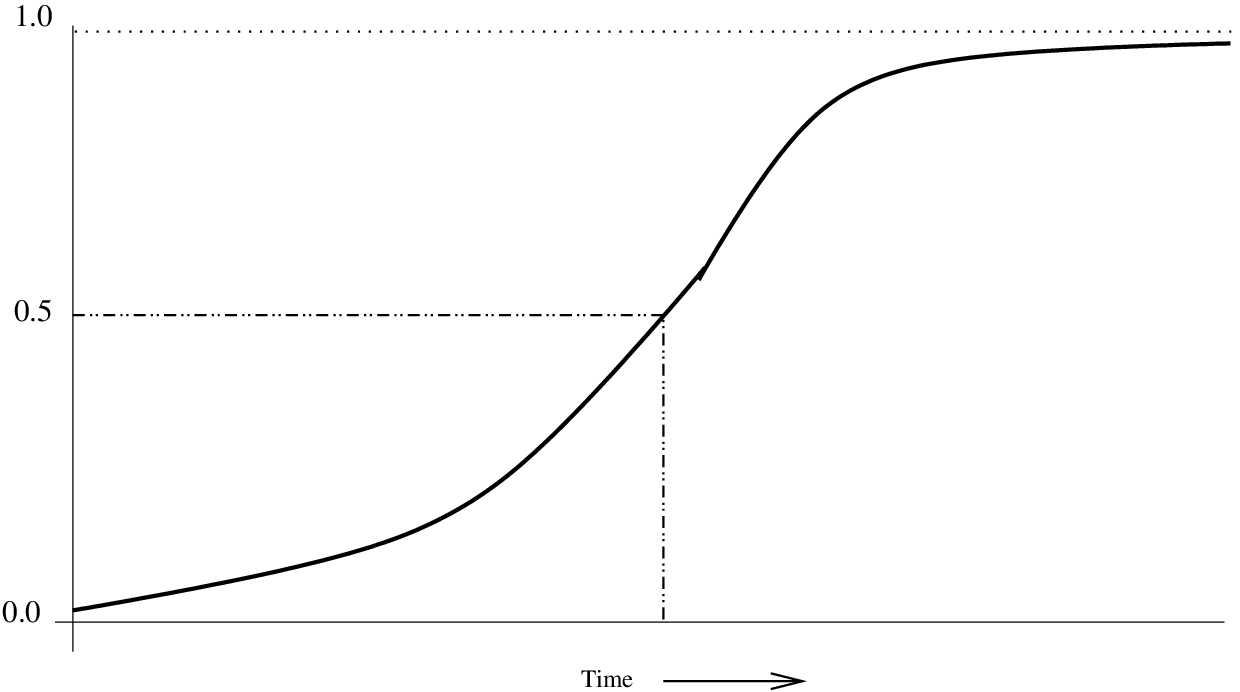}

\caption{\label{fig:Truncated-incubation-period.}Truncated incubation period.
The idea of truncated cumulative distribution of the incubation period
is plotted. After a certain time duration, there will not be any gain
due to therapy. Median incubation period is represented by the line
cutting the curve at 0.5, corresponding to the Y-axis. }

\label{fig-trun}
\end{figure}

The improved models that address the impacts of anti-retroviral therapy,
protease inhibitors and combination of drugs presented in section
1 seem useful in understanding the dynamics of variables for individuals
with the full blown disease for no-drug, \emph{drug1, drug2} and \emph{drug3,
i.e} $D_{z_{0}},D_{z_{1}},D_{z_{2}}$and $D_{z_{3}}.$ Using the methodology
in sections 2 to 4, (despite being lengthy), one can able to estimate
the parameters for the incubation period for each drug type, by the
deconvolution method. We have demonstrated this method for three types
of drugs, and one can obtain $\bold B$ for as many drugs as possible
from the formulas for $n-$types of drugs in section 3. So far, there
is no evidence of drugs being useful in avoiding contracting the disease.
Drugs may be useful for avoiding opportunistic infections for some
specific periods of time. Eventually, an individual will succumb to
AIDS, whether or not that individual takes drugs (which is also demonstrated
in the truncation effect in Figure \ref{fig-trun}). The truncation
effect formulas can be used to obtain the parameter set (say, $\bold B^{T}$),
but we did not demonstrate this numerically. There were other type
of methods for obtaining incubation periods (see \cite{RaoASRS2005-2}
when data is censored and see \cite{Rao&Hira2003} when data is from
hospital based cohort)

We did not introduce intracellular delay that might arise due drug
interventions. There are not many quantitative results available on
the relationship between the dose of a drug and the resultant delay
in the development of the disease. Suppose $s_{1},s_{2},s_{3}...,s_{k}$
are $k$ levels of doses of a single drug, and $\tau_{1},\tau_{2},\tau_{3},...,\tau_{k}$
are the respective delays obtained in producing a new infected cell.
Then we can write the relation $R^{2}\left(s,\tau\right)$ between
$s$ and $\tau$ as 

\begin{eqnarray*}
\frac{\left\{ \sum_{i=1}^{k}\left(s_{i}-\overline{s}\right)\left(\tau_{i}-\overline{\tau}\right)\right\} ^{2}}{\left\{ \sum_{i=1}^{k}\left(s_{i}-\overline{s}\right)\right\} ^{2}\left\{ \sum_{i=1}^{k}\left(\tau_{i}-\overline{\tau}\right)\right\} ^{2}}
\end{eqnarray*}

$R^{2}\left(s,\tau\right)$ is called the correlation coefficient
of dose-delay. $\overline{s}$ is the mean dose-level and $\overline{\tau}$
is the mean delay. This experiment can be conducted for various doses
$s_{ij}$ (say) for drug type $j=1,2,3...n$. Each drug will produce
a delay depending upon the dose level. From this, the average delay
can be statistically compared to understand the mean dose effect due
to a particular drug, and hence the drug efficacy. However, this does
not give dynamics over the time period, but it is very useful in preparing
the baseline parameters for simulation studies, and also for the models
explained in sections 1, 2 and 5. There might be a possibility of
exploring the impact of delay in the conditional probabilities expressed
in this work.

Our work may be interesting for people working on developing computational
techniques for solving integro-differential equations, algorithms
to solve convolution type equations in epidemiology, and EM-type algorithms.
The age-structure analysis presented is more complicated than analysis
presented for the non-age structured populations, and we provide a
new kind of analysis for the incubation period. When reported disease
cases and densities of the infection are available for a period of
several years in the population, then this kind of analysis offers
a reliable method to estimate the incubation period distribution. 

\pagebreak

\pagebreak

\section{\textbf{Appendix I : Conditional probabilities for generalized multiple
drug impact}}

Here we derive expressions for conditional probabilities when several
drugs are available, and the incubation period is non-monotonic. When
such a situation arises there will be several combinations of orders
of $Zs.$ We take one such situation and write corresponding $Ls$
for the purpose of demonstration.

Suppose $Z_{0}<...<Z_{k}=...=Z_{k+n+1}<...<Z_{N}.$ Let us divide
this into the following two inequalities and an equality: $Z_{0}<...<Z_{k}$
, $Z_{k+1}=...=Z_{k+n}$ and $Z_{k+n+1}<...<Z_{N}.$ If we consider
the first and third inequalities, then 

\begin{eqnarray*}
\mathcal{D}\left(\bold A,\bold B/U_{N_{k}}\right) & = & \int_{0}^{U_{N_{0}}}h(t/A_{N_{0}})G(t-s/B_{N_{0}})ds+\\
\\
 &  & \int_{U_{N_{0}}}^{U_{N_{1}}}h(t/A_{N_{1}})G(t-s/B_{N_{1}})ds\\
\\
 &  & \cdots+\int_{U_{N_{k-1}}}^{U_{N_{k}}}h(t/A_{N_{k}})G(t-s/B_{N_{k}})ds
\end{eqnarray*}

\begin{eqnarray*}
\mathcal{D}\left(\bold A,\bold B/U_{N_{N}}\right) & = & \int_{0}^{U_{N_{n+k+1}}}h(t/A_{N_{n+k+1}})G(t-s/B_{N_{n+k+1}})ds+\\
\\
 &  & \int_{U_{N_{0}}}^{U_{N_{n+k+2}}}h(t/A_{N_{n+k+2}})G(t-s/B_{N_{n+k+2}})ds\\
\\
 &  & \cdots+\int_{U_{N_{N-1}}}^{U_{N_{N}}}h(t/A_{N_{N}})G(t-s/B_{N_{N}})ds
\end{eqnarray*}
We can express $\left\{ P\left(E_{N_{\theta}}/E\right)\right\} _{\theta=0}^{\theta=k}$
and $\left\{ P\left(E_{N_{\theta}}/E\right)\right\} _{\theta=n+k+1}^{\theta=N}$
and the corresponding $\left\{ L_{N_{\theta}}\right\} _{\theta}$
as shown in the section 3. Then $L_{N_{\theta}}$ is maximized for
the set $\left[A_{\theta},B_{\theta}\right]$. We obtain $N-n-k$
sets of $\left[\bold A,\bold B\right]$ values, and the corresponding
likelihood functions $\left\{ L_{N_{\theta}}\right\} _{\theta=0}^{k}$
and $\left\{ L_{N_{\theta}}\right\} _{\theta=n+k+1}^{N}$

\pagebreak

\section{\textbf{Appendix II: Truncated incubation period}}

Suppose there is an upper bound for the impact of drugs on the incubation
period; that is, the incubation period cannot be increased after a
certain time point after the drug use. Then the likelihood equations
explained in section 4 would change accordingly. There was an attempt
earlier to truncate the incubation period with the help of the truncated
Weibull distribution \cite{RaoASRS2005-1}. They have not seen the
impact of drugs using such functions. If $Z,$ the length of the incubation
period, and if $Z_{c}$ is the truncation point, then $G(Z)=1-\exp\left\{ -\left(\frac{z}{\delta_{1}}\right)^{\delta_{2}}\right\} ,$
for $0<Z<Z_{c},$ and $G(\mathcal{Z})=1-\exp\left\{ -\left(\frac{z}{\delta_{1}}\right)^{\delta_{2}}\right\} $
$\exp\left\{ -\left(\frac{\delta_{2}}{\delta_{1}}\right)\left(\frac{t_{c}}{\delta_{1}}\right)^{\left(\delta_{2}-1\right)\left(z-z_{c}\right)}\right\} $,
for $Z\geq Z_{c}.$ Here, $\delta_{1},\delta_{2}$ are scale and shape
parameters. One can construct a likelihood function for each drug
type using such functions as follows:

\begin{eqnarray}
L\left(\bold A,\bold B/P_{j}\right) & = & L_{<Z_{c}}+L_{\geq Z_{c}},\label{A1}
\end{eqnarray}
 where

\begin{eqnarray*}
L_{<Z_{c}} & = & \prod_{j}b_{1}(j)b_{2}(j)-\prod_{j}b_{1}(j-1)b_{2}(j)\\
\\
L_{\geq Z_{c}} & = & \prod_{j}b_{1}^{t}(j)b_{2}^{t}(j)-\prod_{j}b_{1}^{t}(j-1)b_{2}^{t}(j)
\end{eqnarray*}

and

\begin{eqnarray*}
b_{1}(j) & = & \left\{ \int_{0}^{u_{j}}e^{\alpha_{1}s^{2}+\alpha_{2}s+\alpha_{3}}\left\{ 1-\exp\left\{ -\left(\frac{t}{\delta_{1}}\right)^{\delta_{2}}\right\} \right\} ds\right\} ^{T_{j}}\\
\\
b_{1}(j-1) & = & \left\{ \int_{0}^{u_{j-1}}e^{\alpha_{1}s^{2}+\alpha_{2}s+\alpha_{3}}\left\{ 1-\exp\left\{ -\left(\frac{t}{\delta_{1}}\right)^{\delta_{2}}\right\} \right\} ds\right\} ^{T_{j}}\\
\\
b_{2}(j) & = & \left\{ \int_{0}^{u_{n}}e^{\alpha_{1}s^{2}+\alpha_{2}s+\alpha_{3}}\left\{ 1-\exp\left\{ -\left(\frac{t}{\delta_{1}}\right)^{\delta_{2}}\right\} \right\} ds\right\} ^{-T_{j}}\\
\\
b_{1}^{t}(j) & = & \left[\int_{0}^{u_{j}}e^{\alpha_{1}s^{2}+\alpha_{2}s+\alpha_{3}}1-\exp\left\{ -\left(\frac{z}{\delta_{1}}\right)^{\delta_{2}}\right\} \times\right.\\
\\
 &  & \qquad\qquad\qquad\left.\exp\left\{ -\left(\frac{\delta_{2}}{\delta_{1}}\right)\left(\frac{t_{c}}{\delta_{1}}\right)^{\left(\delta_{2}-1\right)\left(z-z_{c}\right)}\right\} ds\right]^{T_{j}}\\
\\
b_{1}^{t}(j-1) & = & \left[\int_{0}^{u_{j-1}}e^{\alpha_{1}s^{2}+\alpha_{2}s+\alpha_{3}}1-\exp\left\{ -\left(\frac{z}{\delta_{1}}\right)^{\delta_{2}}\right\} \times\right.\\
\\
 &  & \qquad\qquad\qquad\left.\exp\left\{ -\left(\frac{\delta_{2}}{\delta_{1}}\right)\left(\frac{t_{c}}{\delta_{1}}\right)^{\left(\delta_{2}-1\right)\left(z-z_{c}\right)}\right\} ds\right]^{T_{j}}\\
b_{2}^{t}(j) & = & \left[\int_{0}^{u_{n}}e^{\alpha_{1}s^{2}+\alpha_{2}s+\alpha_{3}}1-\exp\left\{ -\left(\frac{z}{\delta_{1}}\right)^{\delta_{2}}\right\} \times\right.\\
\\
 &  & \qquad\qquad\qquad\left.\exp\left\{ -\left(\frac{\delta_{2}}{\delta_{1}}\right)\left(\frac{t_{c}}{\delta_{1}}\right)^{\left(\delta_{2}-1\right)\left(z-z_{c}\right)}\right\} ds\right].^{-T_{j}}
\end{eqnarray*}

For each drug type, an expression of the type in (\ref{A1}) can be
derived. Despite the assumption on truncation as mentioned above,
the incubation period could vary according to the type of the drug. 

\pagebreak

\section{\textbf{APPENDIX III : Parameters}}

\vspace{4cm}

\begin{table}
\caption{Parameters}

\includegraphics[scale=0.75]{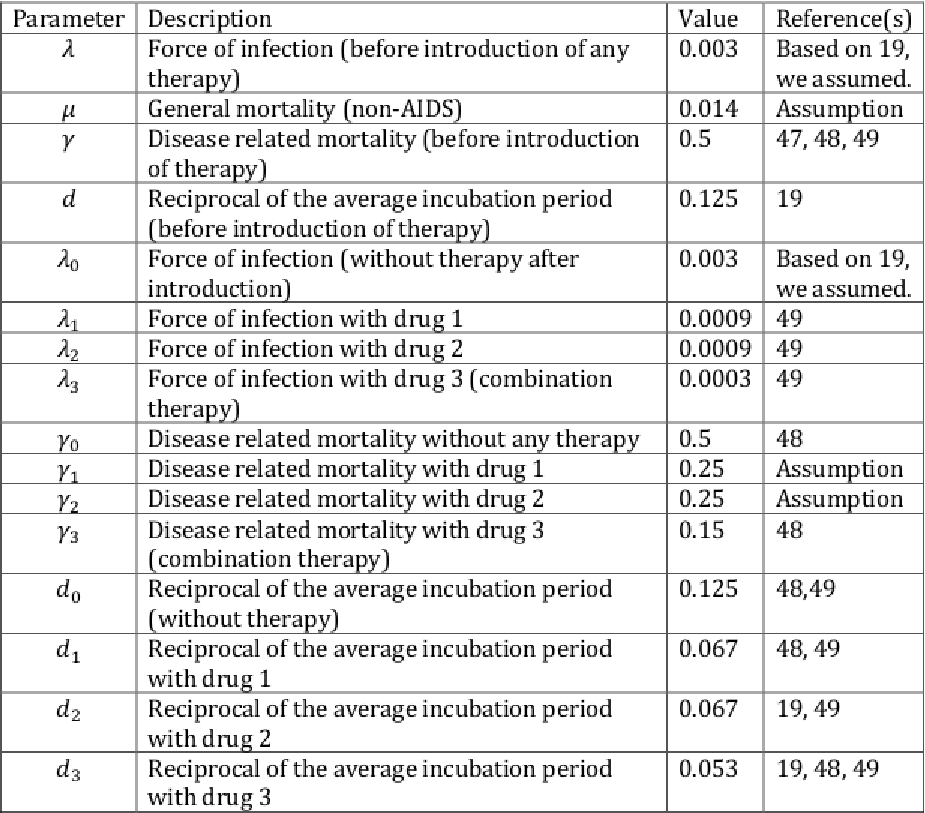}

\end{table}

\pagebreak

\section{\textbf{\textup{APPENDIX IV: FIGURES}}}

\begin{figure}
(a)

\includegraphics[scale=0.8]{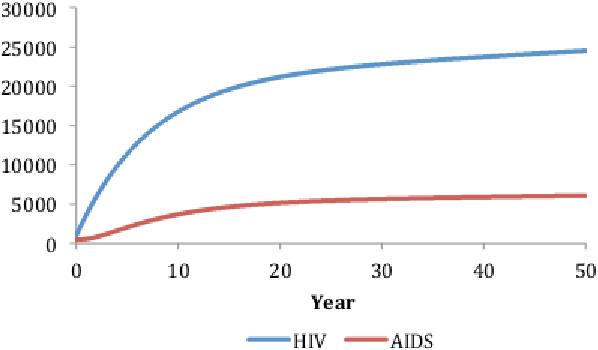}

(b)

\includegraphics[scale=0.8]{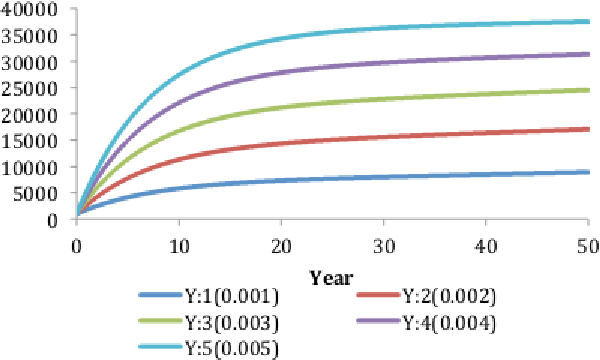}

(c)

\includegraphics[scale=0.8]{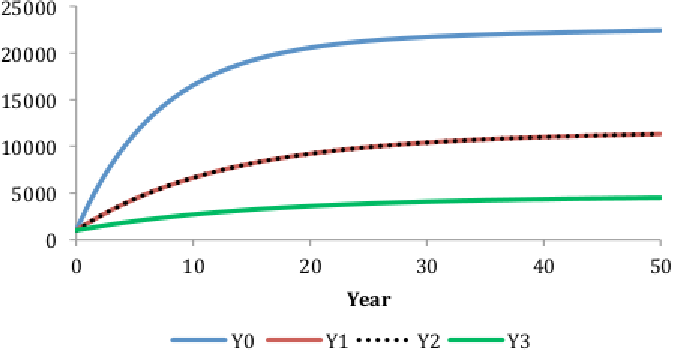}

\label{Appendix IV - Fig1}\caption{(a) Number of HIV and AIDS (before therapy), (b) Number of HIV infected
(before therapy), (c) Number of HIV (after therapy)}

\end{figure}

In this section using the parameters in the Appendix III, output of
the models for hypothetical population sizes and sensitivity of the
parameters in projecting HIV and AIDS are shown through Figure \ref{Appendix IV - Fig1}.1
to Figure \ref{Fig-last}.5.

\begin{figure}
(a)

\includegraphics[scale=0.8]{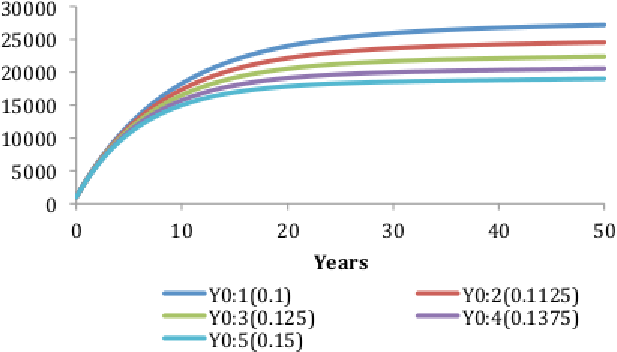}

(b)

\includegraphics[scale=0.8]{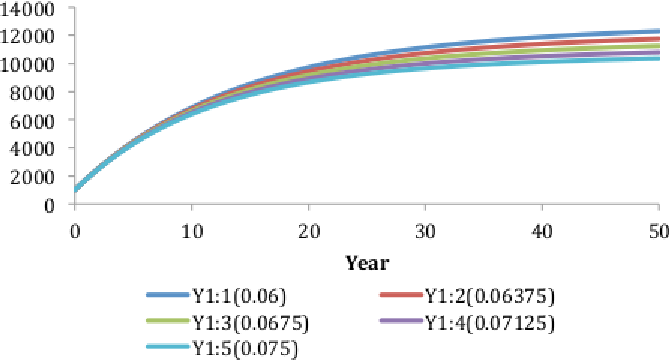}

(c)

\includegraphics[scale=0.8]{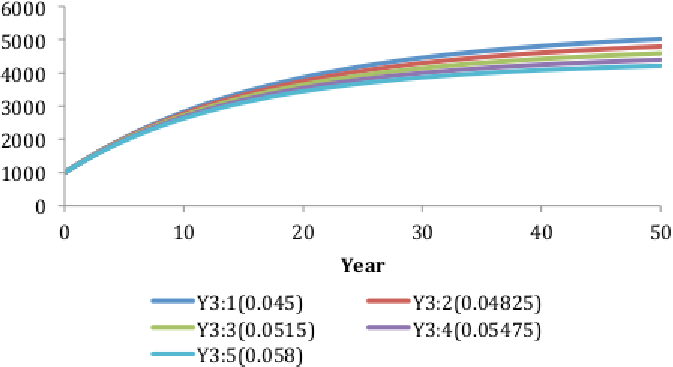}

\caption{(a) Sensitivity of $d_{0}$ on $Y_{0}$, (b) Sensitivity of $d_{1}$
and $d_{2}$ on $Y_{1}$ and $Y_{2}$, (c) Sensitivity of $d_{3}$
on $Y_{3}$.}

\end{figure}

\begin{figure}
(a)

\includegraphics[scale=0.8]{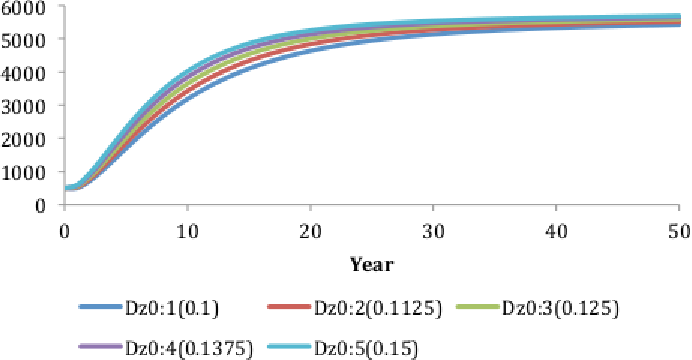}

(b)

\includegraphics[scale=0.8]{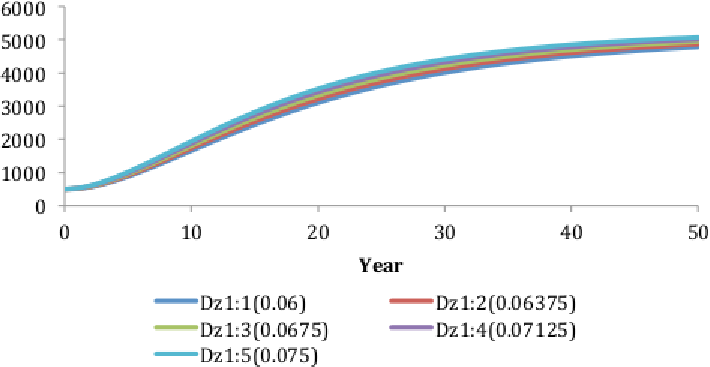}

(c)

\includegraphics[scale=0.8]{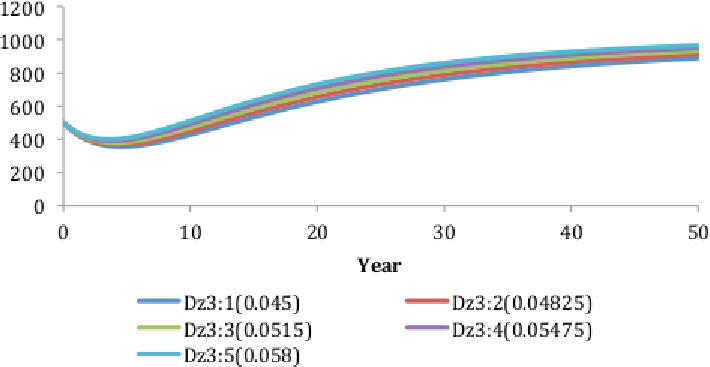}

\caption{(a) Sensitivity of $d_{0}$ on $D_{z0}$, (b) Sensitivity of $d_{1}$
and $d_{2}$ on $D_{z1}$ and $D_{z2}$, (c) Sensitivity of $d_{3}$
on $D_{z3}$.}

\end{figure}

\begin{figure}
(a)

\includegraphics[scale=0.8]{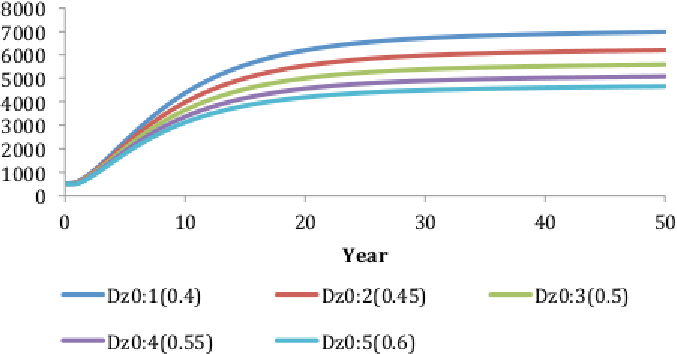}

(b)

\includegraphics[scale=0.8]{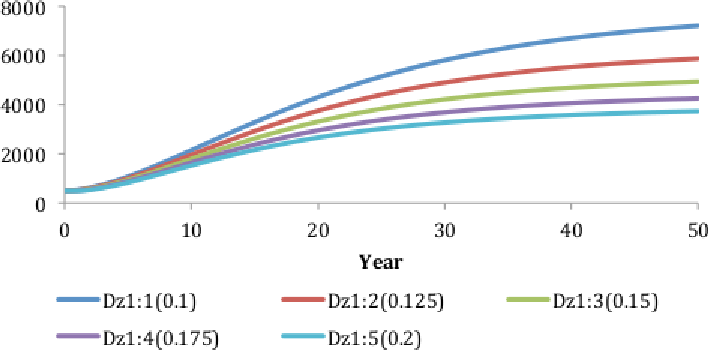}

(c)

\includegraphics[scale=0.8]{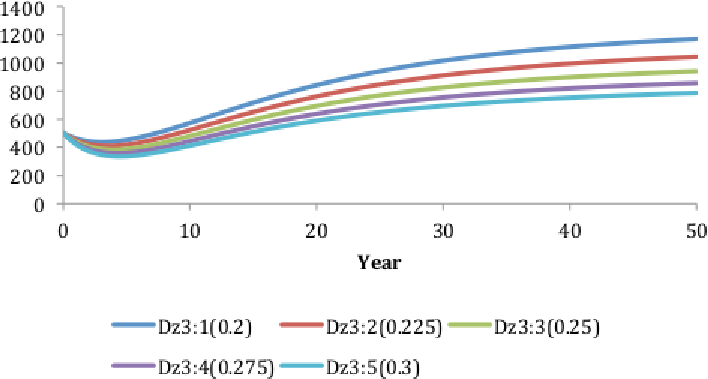}

\caption{(a) Sensitivity of $\gamma_{0}$ on $D_{z0}$, (b) Sensitivity of
$\gamma_{1}$ and $\gamma_{2}$ on $D_{z1}$ and $D_{z2}$, (c) Sensitivity
of $\gamma_{3}$ on $D_{z3}$.}

\end{figure}

\begin{figure}
(a)

\includegraphics[scale=0.8]{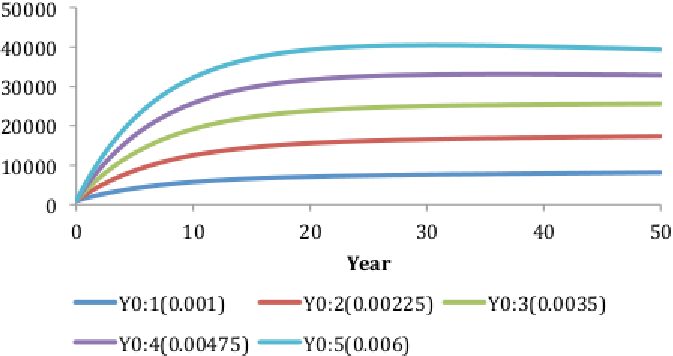}

(b)

\includegraphics[scale=0.8]{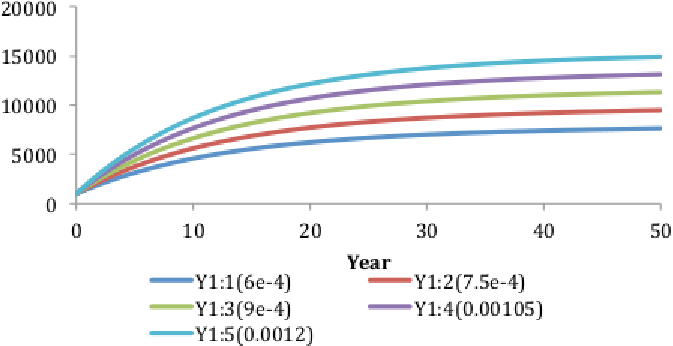}

(c)

\includegraphics[scale=0.8]{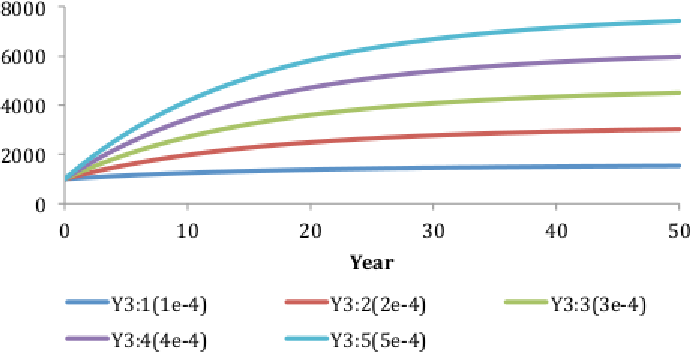}

\label{Fig-last}\caption{(a) Sensitivity of $\lambda_{0}$ on $D_{z0}$, (b) Sensitivity of
$\lambda_{1}$ and $\lambda_{2}$ on $D_{z1}$ and $D_{z2}$, (c)
Sensitivity of $\lambda_{3}$ on $D_{z3}$.}

\end{figure}

\end{document}